\documentclass[superscriptaddress,longbibliography,onecolumn,preprint]{revtex4-1}

\usepackage{xcolor}
\usepackage{amsmath,amssymb,graphicx}
\usepackage{graphicx}

\usepackage{epstopdf}
\usepackage{mathtools}
\usepackage{upgreek}
\usepackage{bm}
\usepackage{color}
\usepackage{braket}
\usepackage{textgreek}
\usepackage{soul}
\usepackage{lipsum}
\usepackage{lineno}
\setstcolor{red}
\usepackage{float} 
\usepackage{environ}
\usepackage{siunitx}
\usepackage{hyphenat}
\usepackage{siunitx}
\begin{document}
\title{High-dimensional topological photonic entanglement}
\author{M.~Javad~Zakeri}
\affiliation{CREOL, The College of Optics and Photonics, University of Central Florida, Orlando, Florida 32816, USA}

\author{Armando Perez-Leija}
\affiliation{Department of Electrical and Computer Engineering, Saint Louis University, St. Louis, Missouri 63103, USA}

\author{and Andrea~Blanco-Redondo}
\email{andrea.blancoredondo@ucf.edu}
\affiliation{CREOL, The College of Optics and Photonics, University of Central Florida, Orlando, Florida 32816, USA}

\date{\today}

\begin{abstract}

The robust generation and manipulation of high-dimensional quantum states lies at the heart of modern quantum computation. The use of topology to resiliently encode and transport quantum information has been widely investigated in condensed matter and has recently penetrated quantum photonics. However, a route to scale up to a large number of entangled topological photonic modes had been missing. Here, we propose and experimentally demonstrate a method to generate high-dimensional topological photonic entanglement. Our platform relies on carefully designed silicon photonic waveguide topological superlattices, which support nonlinear generation of energy-time entangled photon pairs on a superposition of multiple topological modes. Our measurements and theoretical analysis reveal entanglement of up to five topological modes with resilience to nanofabrication imperfections. This study, at the intersection of nonlinear integrated photonics, quantum information, and topology, opens a research avenue toward scalable, fault‑tolerant quantum photonic states.

\end{abstract}

\maketitle

\section{Introduction}
High-dimensional entanglement, a multifaceted phenomenon involving multiple particles in multilevel quantum systems, can be used to increase quantum information capacity and resistance to noise \cite{erhard2020advances}. Physical platforms that allow encoding quantum information in high dimensions range from superconducting phase qudits \cite{neeley2009emulation} to trapped ions \cite{senko2015realization}, all the way through defects in solid state systems \cite{choi2017observation} and photonic settings \cite{malik2016multi}. Among these, photonic systems offer the highest level of flexibility by supporting entanglement of multiple photons and different degrees of freedom, such as the traditional polarization, path, and frequency degrees of freedom, and the more recently demonstrated topology degree of freedom \cite{blanco2018topological, mittal2018topological, wang2019topologically, WANG2022100003, doyle2022biphoton,dai2022topologically}. 

Topological photonics exploits global band-structure invariants to steer light in ways that are inherently immune to certain types of disorder and defects \cite{Lu2014TopologicalPhotonics, Ozawa2019TopologicalPhotonics, price2022roadmap}.  
In analogy to electronic edge transport in condensed matter topological insulators \cite{vonKlitzing1980quantized, Thouless1982,kane2005quantum}, photonic topological insulators route electromagnetic waves along their boundaries with remarkable resilience \cite{wang2009observation, Rechtsman2013PhotonicFloquet, Hafezi2013ImagingTopological, Khanikaev2013PhotonicTopological}.  
Extending these ideas to the quantum regime, quantum topological photonics leverages protected edge or interface modes to generate, transport, and process non-classical states of light with reduced sensitivity to fabrication errors \cite{blanco2019topological,Hashemi2025,JalaliMehrabad2023,Gao2024}. 

Of particular importance is understanding the role that topology can play in protecting the inherently fragile entanglement of multiphoton quantum states \cite{Rechtsman2016TopologicalProtection, Mittal2016TopologicallyRobust}. Early experimental demonstrations involved spontaneous four-wave mixing (SFWM) generation of photon pairs in a single topological mode, revealing that features of biphoton correlations generated and transported in a topological mode showed increased resilience to certain types of disorder \cite{blanco-redondo2016cleo,blanco2018topological,mittal2018topological, WANG2022100003}. Subsequently, this methodology was expanded to investigate the resilience of multiphoton path entanglement of two-topological modes \cite{wang2019topologically,dai2022topologically} by splitting a common pump between two topological modes. This method to create two-path entanglement, commonly used in integrated quantum photonics \cite{silverstone2014chip}, is, unfortunately, not scalable. A paradigm for entangling a number of topological modes larger than two has remained, hindering the scalability of this approach to high-dimensional entanglement, deemed crucial for quantum computing and quantum information science \cite{aghaee2025scaling}.

Here, we propose and experimentally demonstrate a method to generate biphoton entanglement of multiple topological modes. Our approach relies on the use of multiband topological superlattices \cite{PhysRevA.98.043838, Wang2021TopologicalSuperlattice} that support a large number of co-localized topological interface modes. By exciting a linear superposition of interface modes in carefully engineered silicon photonic topological superlattices, we leverage the inherently high optical nonlinearity of the silicon waveguides to generate biphotons in entangled superpositions of three, four, and five topological modes. Theoretical predictions and quantum correlation measurements in numerous fabricated devices reveal that these entangled states retain their main features despite the inevitable variability of nanofabrication. These robust multimode entangled states provide a scalable route to investigate the role of topology in fault-tolerant quantum photonic circuits and resilient quantum communication links.

\section{Results}
\subsection{A platform for scalable multimode topological photonic entanglement}

\begin{figure}[!htbp]
\centering
  \includegraphics[width=0.5\linewidth]{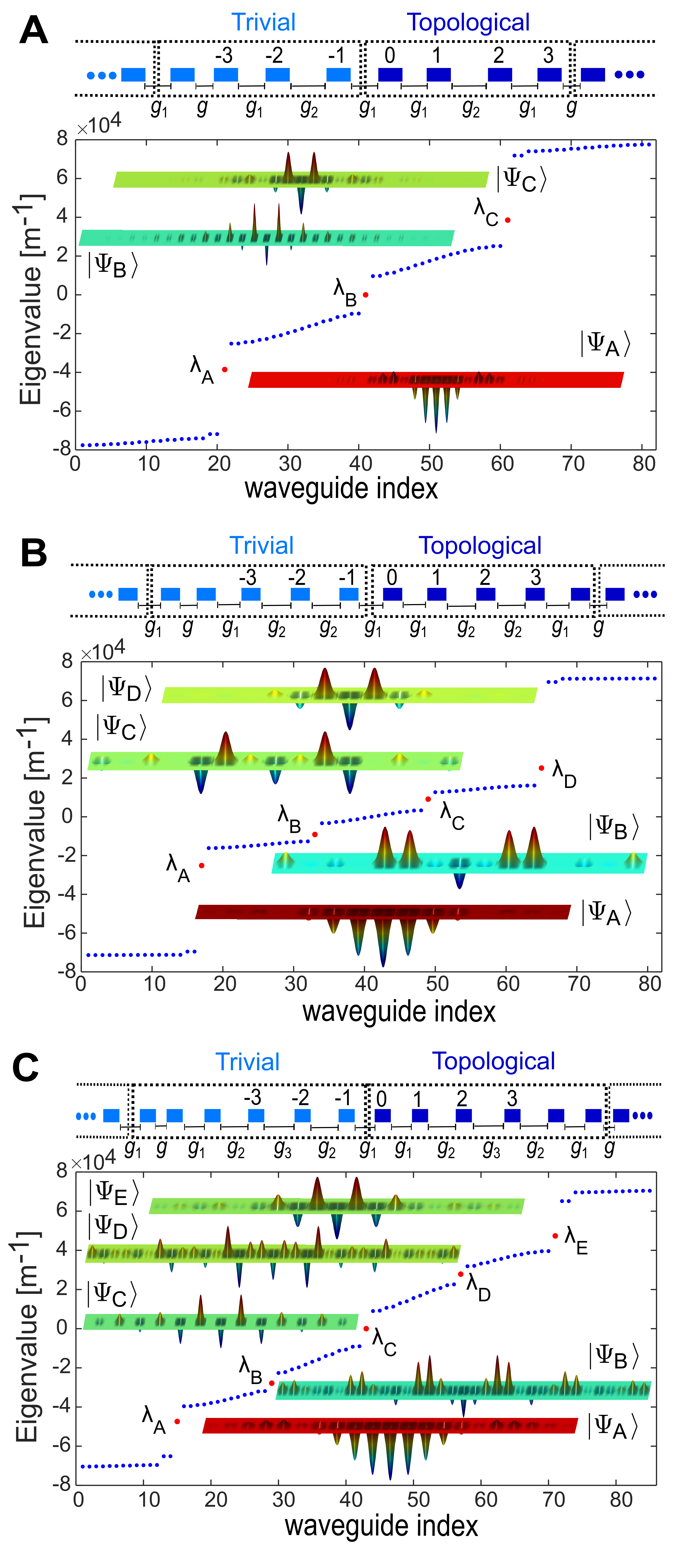}
    \caption{
    Eigenspectrum and mode profiles of the topological superlattice waveguide array (A)--(C) corresponding to supperlattices \(J=4,5\), and 6. In (A) $g_1=$235 nm, $g_2=$261 nm, $g=$120 nm, and $N=$81; in (B) $g_1=$282 nm, $g_2=$330 nm, $g=$125 nm and $N=81$; and in (C) $g_1=$225 nm, $g_2=$230 nm, $g_3=$255 nm, $g=$ 140 nm, and $N=85$. All waveguides are made of silicon (refractive index $n_{\text{Si}}=3.48$ at $1550$ nm wavelength) on a silicon dioxide substrate and have a width of $w = 500$ nm, a height $h = 220$ nm,  and a length $L = 500\mu$m). The cladding is air.
}
    \label{Fig1}
\end{figure}

Silicon photonics provides the intrinsic stability, high precision, and dense integration needed for quantum applications \cite{silverstone2016silicon, silverstone2014chip}. Hence, the core of our experiments is a set of silicon photonic waveguide arrays, each one hosting two interfacing superlattices, one topological and one nontopological (henceforth trivial). Specifically, we implemented three types of arrays with unit-cells formed by J=4,5,6 waveguides, as shown in the insets of Fig. \ref{Fig1}(A--C), respectively. These superlattices are endowed with $J-1$ topological band gaps if they fulfill two main criteria \cite{PhysRevA.98.043838}: the unit cell structure preserves inversion symmetry and the intercell coupling is greater than a particular value at which the largest band gap in the system closes (see supplementary text S1). Following these premises, we carefully engineered the intracell and intercell coupling strengths in the silicon waveguide arrays to create topological-trivial transitions at their center. The theoretically computed spectra for these configurations shown in Fig. 1 unveil the existence of three, four, and five topological gaps inhabited by three, four, and five interface states, respectively. Even though each topological mode resides in different topological gap, their corresponding amplitude distributions coexist spatially localized at the boundary between the trivial and topological superlattices. In contrast, the amplitude of the bulk modes, whose energy is represented by the black dots in the eigenspectra plots, is spread across the lattice. The size of the multiple bandgaps and the localization of the interface states can be engineered by tailoring the intercell and intracell couplings (see supplementary text S2).

To generate entanglement across the multiple co-localized topological modes, we resort to degenerate four-wave mixing (DFWM) intrinsically occurring in the silicon waveguides, which have an inherently high nonlinear parameter of $\gamma = 120~\mathrm{W}^{-1}\,\mathrm{m}^{-1}$. DFWM is a $\chi^{(3)}$ nonlinear optical process in which two pump photons at frequency $\omega_p$ give rise to signal and idler photons at frequencies $\omega_s$ and $\omega_i$, respectively. As energy and momentum are preserved in this process, the resulting signal--idler pairs (biphotons) are energy--time entangled.

In our experiments, the pump light-field distribution is generated by exciting the first waveguide within the topological superlattice (waveguide~0 in the insets of Fig.~1(A--C)). Crucially, this single–waveguide excitation produces a beating pump pattern formed by a classical (non\-quantum) linear superposition of all $(J-1)$ topological modes, as shown in Fig.~2(A--C). It should be emphasized that the observed Bloch-like oscillating dynamics in the pump field involve only topological interface modes; no bulk states are excited. Under this pump-excitation condition, spontaneous four-wave mixing (SFWM) probabilistically generates idler--signal biphotons throughout the trivial--topological superlattice boundary. This readily produces the entangled (nonseparable) multi–topological–mode state
\begin{equation}
\ket{\Psi}
=\sum_{m=1}^{J-1} c_{m,m}\,\ket{\Psi^{s}_{m}}\ket{\Psi^{i}_{m}}
\;+\;
\sum_{m=1}^{J-1}\sum_{n=m+1}^{J-1}
c_{m,n}\!\left(\ket{\Psi^{s}_{m}}\ket{\Psi^{i}_{n}}
+\ket{\Psi^{s}_{n}}\ket{\Psi^{i}_{m}}\right),
\label{eq:multi_topo_state}
\end{equation}
where $\ket{\Psi^{s}_{m}}$ and $\ket{\Psi^{i}_{n}}$ denote the states in which the signal and idler photons populate the $m$-th and $n$-th topological modes, respectively. The amplitudes $c_{m,n}$ can be adjusted by tailoring the superlattice design parameters and the pump distribution. Once generated, this state evolves through the system exhibiting Bloch-like oscillations in correlation space. The theoretical pump and biphoton dynamics are illustrated in Fig.~2(A--C) for superlattices supporting 3-5 topological modes, respectively. The biphoton correlations of the generated quantum states are shown at the output of each lattice.

\begin{figure}[htbp]
    \centering
    \includegraphics[width=1\textwidth]{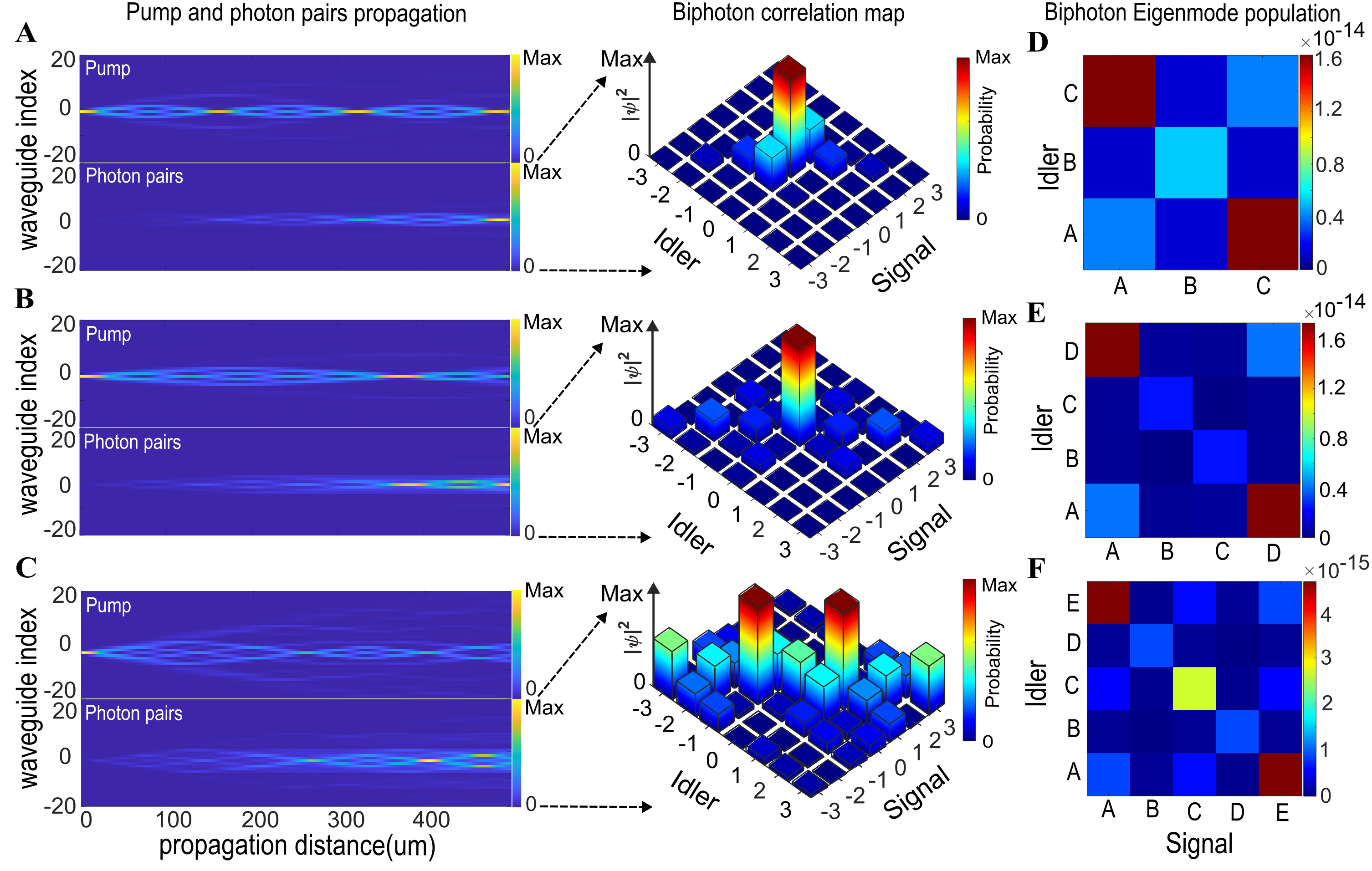 } 
    \caption{Numerical simulation of quantum-state generation in superlattices. (A-C) correspond to J=4, 5, and 6, respectively, and show the pump intensity propagation (top-left), the photon-pair intensity propagation represented by the diagonal elements of the biphoton correlation intensity at each propagation point (bottom-left), and the biphoton correlation maps at the output of the seven central waveguides (right).(D-F) correspond to J=4, 5, 6, respectively, and show the biphoton population matrices showing projections onto the localized eigenmodes.
}
    \label{Fig2}
\end{figure}

To elucidate the topological mode composition of the output state, Figs. \ref{Fig2}(D–F) show the biphoton eigenmode occupancies, computed from the overlap of the output biphoton correlation amplitudes and the biphoton eigenmodes (see supplementary text S4). As pointed out originally in \cite{doyle2022biphoton}, the generation of photon pairs via DFWM in multimode lattices can, in principle, occur between any combination of four eigenmodes, including the bulk modes. However, the probability of any given combination is determined by the overlap between the four mode profiles in the waveguides of interest. Strongly localized modes – in this case topological interface modes – yield higher overlap factors. Therefore, DFWM occurs primarily between the topological interface modes, leading to an entangled state involving only topological modes (see Movies S1, S2, and S3).

The weights of each mode in the entangled state are highly dependent on the degree of modal localization, where the modes hosted by larger gaps (i.e. more strongly localized) accumulate the larger populations than those in narrower gaps. This is illustrated, for example, by the higher weights of the modal combinations involving modes A and E in Fig. \ref{Fig2}(F), in contrast with the lower weights for modal combinations involving modes B, C, or D, which are more weakly localized (see supplementary text 5).

Phase-matching considerations also play a key role, since the biphoton eigenmode populations evolve according to the relative phase accumulated by the signal and idler photons. When signal and idler populate the same mode, their phase mismatch depends exclusively on chromatic dispersion, i.e. the fact that photons at different frequencies ($\omega_s$ and $\omega_i$) travel at different speeds, which is a mild effect in the present system. However, when the signal and idler populate different modes, in addition to chromatic dispersion we must consider modal dispersion, since each interface state has its own propagation constant. In this case, the phase mismatch term varies sinusoidally with the propagation length, leading to propagation distances where these mode combinations show enhanced probability and others where they are completely suppressed probability (see Supplementary Text~S6). This explains why, at relatively low propagation distances, such as the one considered here, $L = 500\,\mu\mathrm{m}$, near a point of minimum phase mismatch, we observe a higher weight for mixed-modal terms in the entanglement (e.g., AD/DA in Fig.\ref{Fig2} 2(E)), whereas at longer propagation distances the weight of the non-mixed terms would eventually prevail.

In lattices with mirror symmetry, every eigenmode is either symmetric or antisymmetric with respect to the central lattice interface, and the modal parity is also expected to play a role \cite{doyle2022biphoton}. In the superlattice case studied here, the period of the phase flips in antisymmetric modes varies from mode to mode (Fig. \ref{Fig1}(A-C)), and the overlap term only becomes zero in cases involving an odd number of antisymmetric modes with the same phase-flip period, for instance, modes B and C in Fig. \ref{Fig1}(C) (see supplementary text 7).

Hence, the structure of the generated entangled state, encoded in the topological mode weights contained in the entangled state, can be conveniently engineered by tuning the spectral gaps and the length of the waveguides.

\subsection{Experimental demonstration of high-dimensional topological photonic entanglement}

\begin{figure}[htbp]
    \centering
    \includegraphics[width=1\textwidth]{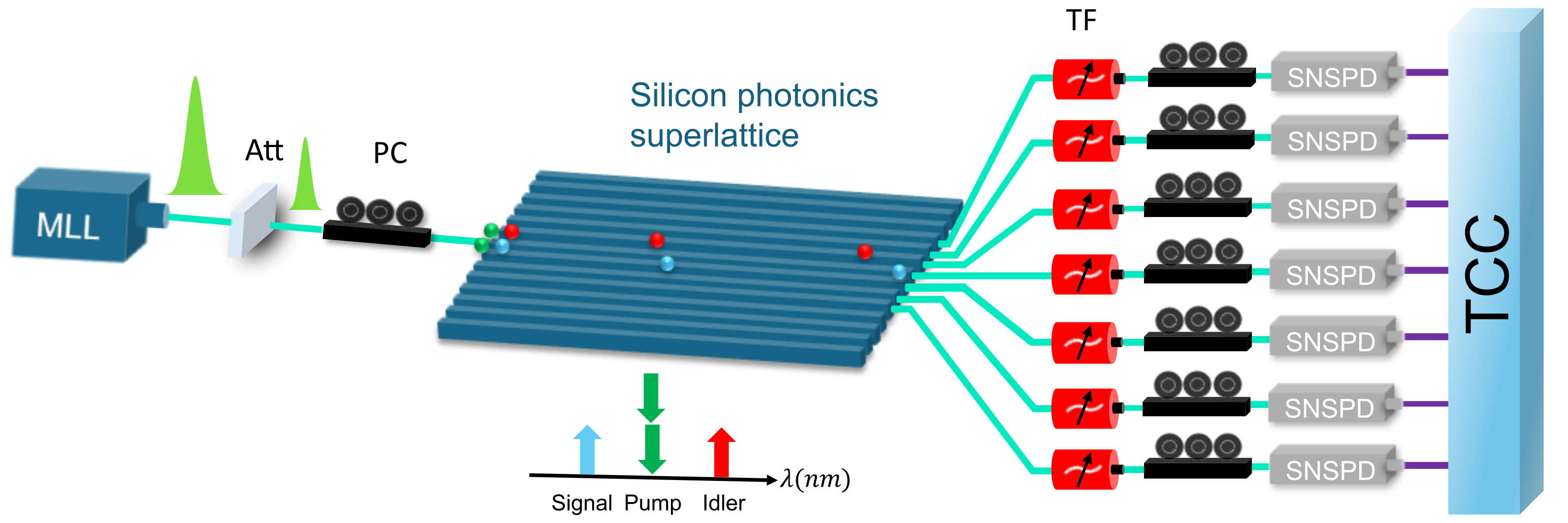} 
     \caption{ Simplified schematics of the experimental setup. A mode locked laser emitting 720 fs pulses at 1550 nm is connected to the chip after passing by an attenuator (Att), polarization controller (PC). The pulses propagate though the silicon photonic superlattices inducing the generation of photon pairs via nonlinear four-wave mixing. These photon pairs are in a superposition of multiple topological modes. At the output of the superlattices, signal and idler photons are spectrally separated by tunable filters (TF) and detected by superconducting nanowire single photon detectors (SNSPDs). The correlations are identified via a time correlation circuit (TCC).}
    \label{Fig3}
\end{figure}

To demonstrate the feasibility and robustness of our approach, we fabricated and characterized four copies of each of the superlattices described above. A schematic of the experimental setup for the generation and measurement of high-dimensional topological entanglement in these lattices is shown in Fig. \ref{Fig3}. Sub-picosecond pump pulses at $1550\,\mathrm{nm}$ are coupled from a fiber into the waveguide at the center of the topological interface of the superlattice using a grating coupler. As they traverse the structure, these high-peak-power pulses drive spontaneous DFWM, generating energy--time entangled photon pairs. At the end of the lattice, we fan out the seven central waveguides and couple light into seven separate single-mode fibers (see Materials and Methods). The output photons are spectrally separated by narrowband filters at $\approx 1545\,\mathrm{nm}$ for the signal and at $\approx 1555\,\mathrm{nm}$ for the idler photons. Subsequently, we detect individual photons using superconducting nanowire single-photon detectors (SNSPDs) and register photon-coincidence events with a high-resolution time--correlation circuit (TCC). This procedure maps the spatial probability intensity of the biphoton wavefunction by showing the pairwise correlations as a function of the lattice site.

\begin{figure}[htbp]
    \centering
    \includegraphics[width=1\textwidth]{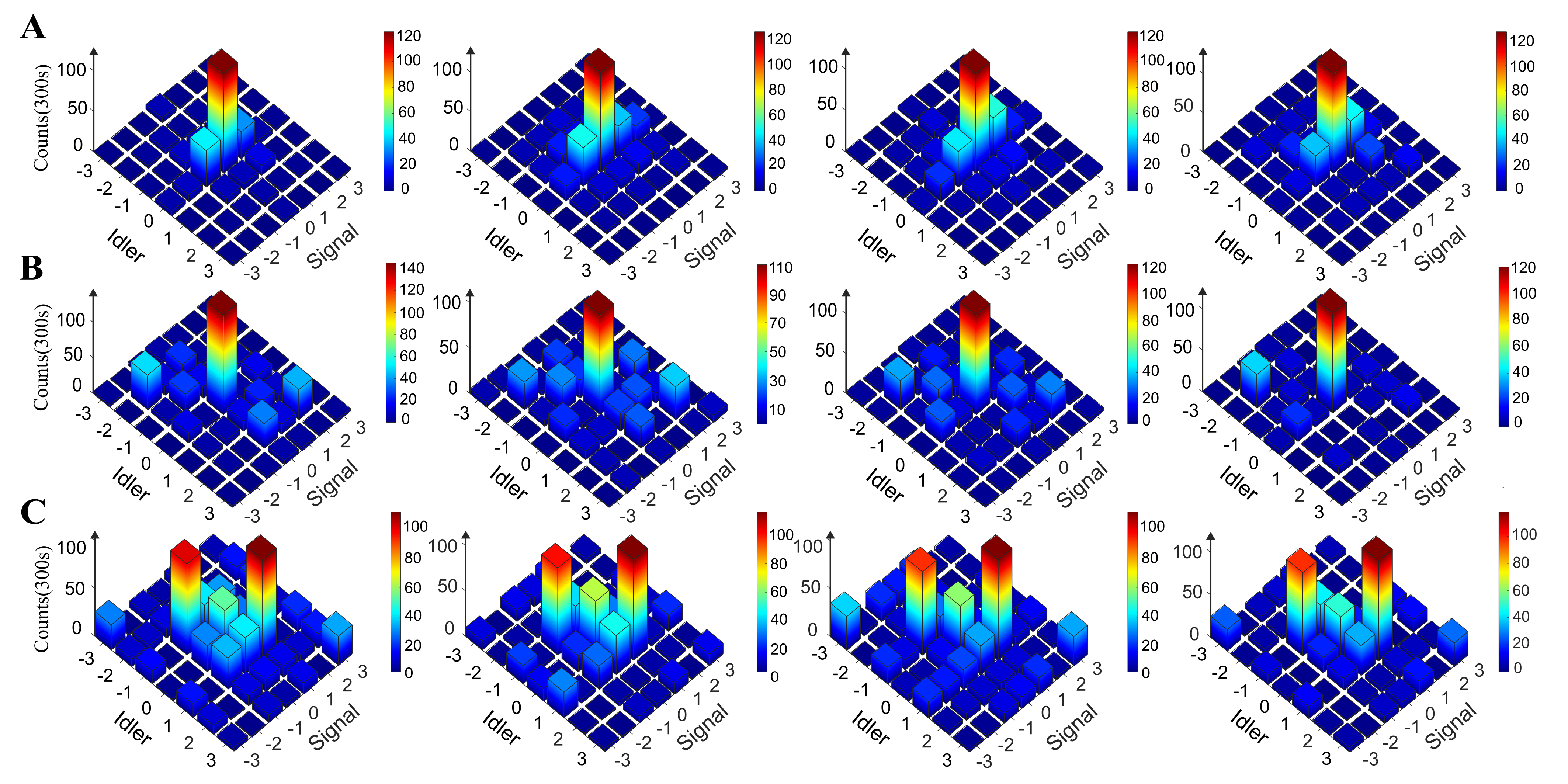} 
    \caption{Measured biphoton correlation maps. (A-C) Show measured correlation maps at the output of seven central waveguides across four different fabricated devices for $J=4$–6, respectively.}
    \label{fig4}
\end{figure}

Figure~\ref{fig4}(A) shows the biphoton correlation maps measured at the output of four fabricated $J=4$ superlattices. Each point in the map is obtained from coincidence counts accumulated over $300\,\mathrm{s}$. Figs \ref{Fig2}(A)--(C) show analogous measurements for superlattices with $J=5$ and $J=6$ waveguides per unit cell, respectively. These measured biphoton correlation maps agree well with the theoretical predictions in Figs \ref{Fig2}(A)--(C) for each superlattice case.
 
The observations in Fig. \ref{fig4} indicate that, in all cases, the biphotons tend to emerge co-localized around the topological interface. Yet, this co-localization effect, showcased by higher photon bunching in the interface waveguide, is more prominent for the superlattices supporting 3 and 4 topological modes as observed in Figs. \ref{fig4}(A) and (B). In contrast, in systems involving 5 topological modes, Fig. \ref{fig4}(C), the output two-photon wavefunctions spread over seven waveguides, featuring the highest probability for biphoton bunching effects in waveguides 1 and -1. These biphoton localization effects are a clear manifestation of the Bloch-like revivals and dynamic localization occurring in correlation space. In other words, for superlattices supporting 3 and 4 topological modes, the biphoton correlation maps present several collapses and revivals of probability at regular intervals along the propagation coordinate. Consequently, photon bunching effects are detected in the output states when a revival point coincides with the superlattice length. Conversely, superlattices supporting 5 interface modes do not exhibit revivals and the two-photon wavefunctions populate several waveguides in the vicinity of the topological boundary.  Nevertheless, in all systems the biphoton wave packets are formed by the associated topological modes, and as such they resist fabrication imperfections.   

Next, we analyze the robustness of the measured entangled states. Although devices 1--4 have identical design parameters for each superlattice case, the inevitable nanofabrication tolerances ($\pm 5~\mathrm{nm}$ for state-of-the-art $100~\mathrm{keV}$ e\mbox{-}beam lithography) lead to physical variations between devices, which affect both the waveguide widths and the gaps between waveguides. Nonetheless, the measured biphoton-correlation intensity maps show remarkable resemblance between devices for the $J=4,5,6$ cases shown in Figs. \ref{fig4}(A)--(C). For a quantitative analysis, we use two distinct measures: the Schmidt number $K$, which quantifies the dimensionality of the entangled states (with $K=1$ for separable states and $K>1$ for entangled states; higher values indicate a larger number of modes involved) \cite{sperling2011schmidt}; and the fidelity with respect to a state in an unperturbed scenario without fabrication disorder, i.e., with respect to the theoretical predictions \cite{Bergamasco2019}.

\begin{figure}[htbp]
    \centering
    \includegraphics[width=0.6\textwidth]{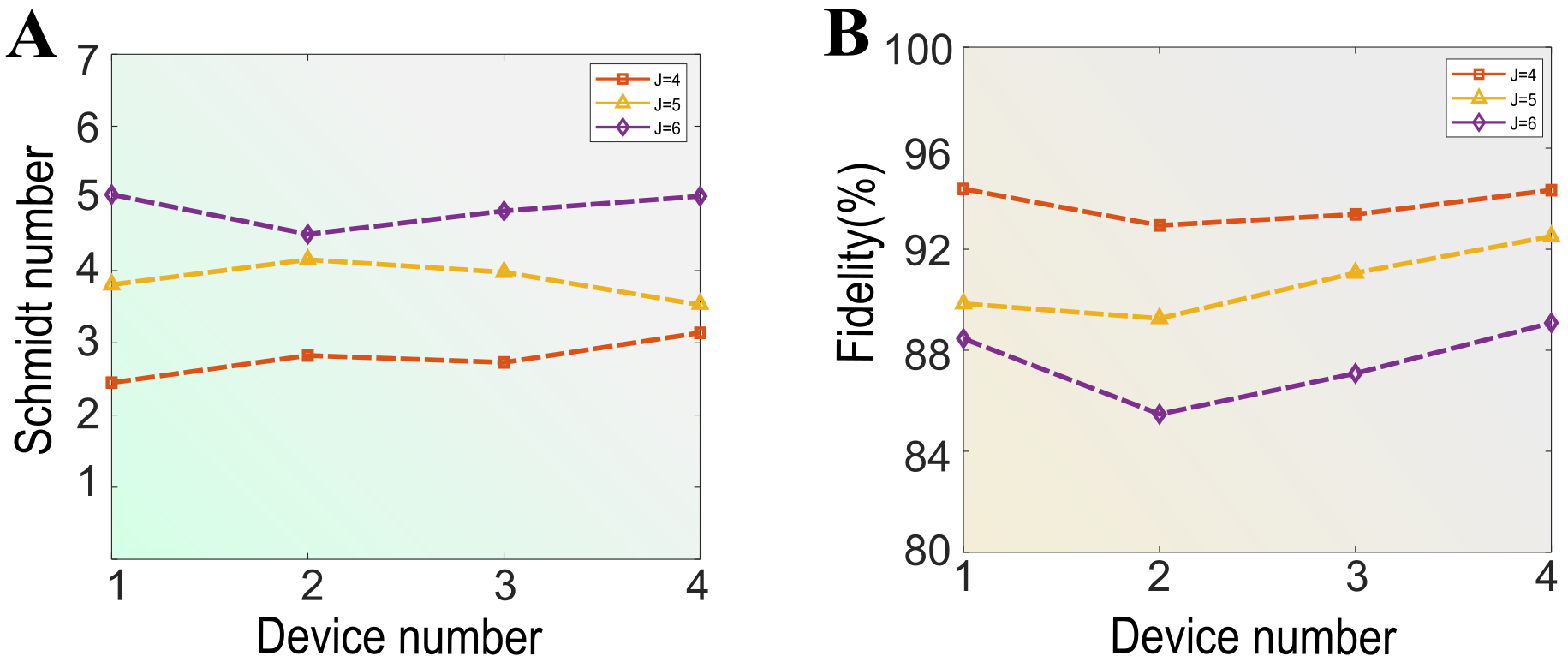} 
    \caption{
   Schmidt number and fidelity of the measured quantum states across devices. (A) Schmidt number versus device index in the topological superlattice platform, for varying \(J\).
   (B) Fidelity versus device index in the same platform for different \(J\).
    }

    \label{Fig5}
\end{figure}

Figure \ref{Fig5}(A) shows the Schmidt number of the state measured at the output of each of the four devices for J = 4, 5, and 6. As J increases, so does the Schmidt number, as expected from the higher number of entangled topological modes. Importantly, the Schmidt number remains relatively unchanged across the four devices for each superlattice case. The fidelity, depicted in Fig. \ref{Fig5}(B), also remains consistently high across devices for each superlattice. The average fidelity of the measured quantum states at the output of the superlattices appears to decrease slightly with J, indicating that the highest-dimensional entangled states are less robust to disorder. This is linked to the decreasing size of the bandgaps as can be appreciated by comparing, for instance, the eigenspectra in Fig. \ref{Fig1}(A) and Fig. \ref{Fig1}(B), and it can possibly be compensated by band gap engineering.

\section{Conclusion}

We proposed and demonstrated a method for the generation of high-dimensional topological photonic entanglement. Our approach, based on the combined use of multiband topological superlattices and nonlinear quantum state generation, represents a versatile platform to explore the interaction of topology and quantum information in multiphoton and multimode systems. Our measurements and analysis confirm that entanglement dimensionality scales predictably with the complexity of the unit cell J, providing a controlled pathway toward larger Hilbert spaces of topologically protected modes, where the entanglement can be shaped by tuning physical parameters of the structure. The similarity of the experimental measurements of the biphoton correlation maps across different devices, in addition to quantitative analysis via Schmidt decomposition and overlap metrics, verifies the resilience of the generated quantum states to nanofabrication tolerances.

Beyond the role of topology, the extraordinarily rich physics available in this relatively simple and accessible platform composed of conventional silicon photonic waveguides, provides a convenient playground to explore other multimode quantum phenomena, such as parity-related entanglement \cite{selim2025selective, roy2025parity}, hyperentanglemet \cite{shaw2025erasure}  and quantum interference \cite{Tambasco2018, Ehrhardt2024}.

At a point when it is becoming increasingly clear that building highly complex, error-resilient quantum states is key to building scalable quantum computing systems that can target useful problems \cite{larsen2025integrated}, we expect our findings to contribute to clarifying the fundamental role that topology can play in protecting high-dimensional quantum information.

\begin{acknowledgments}
A.B.-R. and M.J.Z. are supported by the National Science Foundation (NSF) (award ID 2328993). A P.-L. is supported by MURI grant from Air Force Research Office (programmable systems with non-Hermitian quantum dynamics: FA9550-21-1-0202).
\end{acknowledgments}

\clearpage

\begin{center}
  {\Large\bfseries Supplementary Material for}\\[12pt]
  {\LARGE\bfseries High-dimensional topological photonic entanglement}\\[16pt]
  \large
  
\end{center}

\section*{Materials and methods}

\underline{Fabrication considerations}

The devices were fabricated on 220 nm-thick silicon-on-insulator (SOI) wafers and intentionally left air-clad on the top surface. The SOI platform provides a large refractive index contrast between silicon \((n_{\mathrm{Si}}\approx 3.48)\) and air \((n\approx 1)\), yielding strong modal confinement and enabling bend radii below \(10~\mu\mathrm{m}\). Retaining the air cladding obviates the deposition of a top oxide, which would otherwise reduce the effective index contrast, broaden the band-edge dispersion, and weaken the topological features central to our design. The air interface further mitigates thermally driven phase drift in quantum-optics measurements by eliminating oxide-related absorption pathways.

The layouts were exported to GDSII using a dedicated process design kit (PDK) that enforced key fabrication rules: a minimum trench width (gap) of \(70~\mathrm{nm}\), adiabatic edge tape tips of \(120~\mathrm{nm}\) and a local radius of curvature \(\ge 200~\mathrm{nm}\) at corners to mitigate post-etch sidewall roughness. The patterns were written by high-resolution electron-beam lithography and transferred with a precisely calibrated two-step inductively coupled plasma (ICP) etch. This flow yielded uniform critical dimensions and smooth, near-vertical sidewalls, keeping waveguide and slot widths within tolerance and preserving the intended topological coupling performance.

\subsection*{\textbf{\underline{Coupling Methods}}}

Optical coupling to and from the chip is provided by TE polarized grating couplers (GC) optimized for a launch angle of \SI{12}{\degree} in air and spectrally centered at \SI{1550}{\nano\metre}, with a bandwidth of approximately \(\pm\SI{20}{\nano\metre}\). The grating period and duty cycle are tailored to \SI{12}{\degree}-polished single‑mode fibers, minimizing Fresnel reflections and maintaining near‑normal incidence within the silicon waveguides. Experimentally, each GC exhibits an insertion loss of \(\sim\SI{6.7}{\decibel}\) at \SI{1550}{\nano\metre}, consistent with the NanoSOI PDK characterization data.

To suppress radiation losses from waveguide bends without enlarging the device footprint, access waveguides incorporate adiabatic flares upstream of the first curvature, allowing the fundamental mode to expand gradually. Full vector electromagnetic simulations confirm that residual bend radiation remains at least \SI{10}{\decibel} below the experimental noise floor.

\begin{figure}[H]
    \centering
    \includegraphics[width=0.7\textwidth]{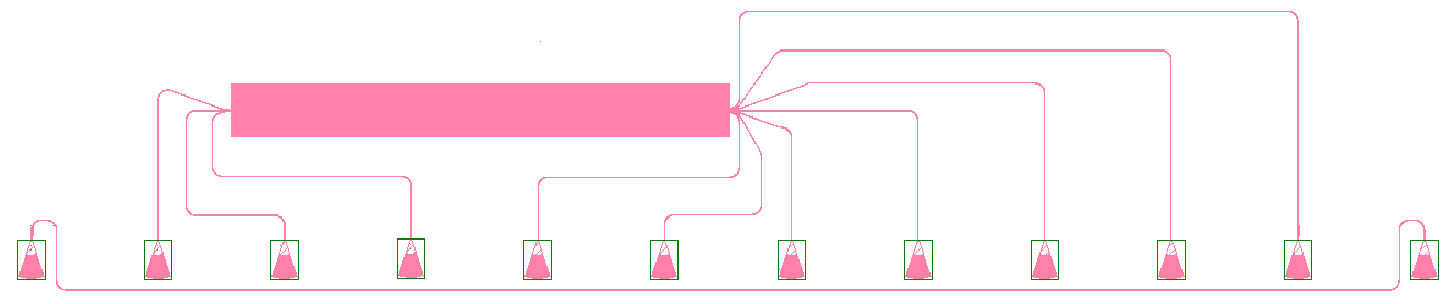}
    \caption{Topological superlattice chip with a 12\hyp{}channel grating coupler array matched to a \(250~\mu\mathrm{m}\)\hyp{}pitch fiber array. Three ports inject the pump, seven collect lattice outputs, and two serve as alignment fiducial and insertion\hyp{}loss references.}
    \label{fig:couplingS}
\end{figure}

All grating couplers are laid out to match a 12\hyp{}channel fiber array with \(250~\mu\mathrm{m}\) pitch (Fig. \ref{fig:couplingS}): three deliver the pump, seven collect lattice outputs, and two serve as alignment fiducials and insertion\hyp{}loss references. Patterning the full array in a single electron\hyp{}beam exposure limits channel\hyp{}to\hyp{}channel placement error to within the tool’s sub\hyp{}\(100~\mathrm{nm}\) overlay tolerance, enabling repeatable, low\hyp{}loss coupling across measurements.

\underline{Propagation Simulations}

To coherently link the static band structure embedded in the signal and idler lattice Hamiltonians with the dynamical evolution of biphoton correlations, we carried out comprehensive numerical simulations of spontaneous four-wave mixing in the engineered superlattices. The biphoton wavefunction \(\psi(z) \in \mathbb{C}^{N \times N}\) evolves along the propagation coordinate \(z\) according to the discrete Schrödinger equation.
\begin{equation}
i\frac{\partial \psi}{\partial z} 
= \psi H_i^* + H_s \psi + \gamma \psi_0 \, \mathrm{diag}\bigl(A_{1,p}^2(z), \dots, A_{N,p}^2(z)\bigr),
\label{biphoton_schrodinger}
\end{equation}
where \(H_s\) and \(H_i\) represent the Hamiltonians of the signal and idler waveguide lattices, respectively, \(\gamma\) is the nonlinear coefficient, and \(A_{n,p}(z)\) denotes the pump amplitude in the \(n^{\mathrm{th}}\) waveguide (see supplementary material in Ref.~\cite{blanco2018topological}). The global biphoton generation efficiency \(\psi_0\) encapsulates the effective nonlinear interaction strength, depending explicitly on waveguide parameters such as the nonlinear refractive index \(n_2\), effective mode area \(A_{\mathrm{eff}}\), central wavelength \(\lambda_0\), and mode overlap integrals. The dynamics described by Eq.~\eqref{biphoton_schrodinger} fundamentally depend on several physical and structural parameters, including the nonlinear coefficient, fabrication-induced disorder, and numerical precision, detailed below.

\paragraph{Nonlinear coefficient \(\gamma\).}
Silicon waveguides exhibit a significant instantaneous Kerr response, characterized by an intensity-dependent refractive index change \(\Delta n = n_{2}I\). At \(\lambda_{0}=1550~\mathrm{nm}\), the Kerr coefficient of silicon is \(n_{2}=6.0\times10^{-18}\,\mathrm{m^{2}/W}\). The waveguide nonlinearity \(\gamma\) is given by
\[
\gamma = \frac{\omega_{0}\,n_{2}}{c\,A_{\mathrm{eff}}},
\]
where, for an effective mode area \(A_{\mathrm{eff}} \approx 0.10~\mu\mathrm{m^{2}}\), we find \(\gamma = 120~\mathrm{W^{-1}m^{-1}}\). Due to the linear dependence of Eq.~\eqref{biphoton_schrodinger} on \(\gamma\), scaling either pump power or interaction length by a factor \(\eta\) proportionally scales the biphoton amplitude by the same factor, provided that phase-matching conditions remain stable and pump depletion negligible.

\paragraph{Disorder modeling.}
Variations introduced by electron-beam lithography and etching steps manifest themselves as random fluctuations in the strengths of the nearest-neighbor coupling. These are captured mathematically by introducing random perturbations:
\[
J_{i-1,i},\,J_{i,i-1}\;\rightarrow\;
J_{i-1,i}(\delta_i),\;J_{i,i-1}(\delta_i),\quad \delta_i\sim\mathcal{N}(1,\sigma^{2}).
\label{eq:disorder}
\]

The dimensionless disorder parameter \(D\), defined as \(\sigma = D\), quantifies the magnitude of these fluctuations. Typical silicon photonic fabrication tolerances correspond to values of \(D\) ranging from a few percent to tens of percent. The model accurately captures the dominant lattice randomness without requiring fitted parameters. Consistently applying the same random factors \(\{\delta_i\}\) in \(H_p\), \(H_s\), and \(H_i\) ensures a physically consistent disorder profile. By performing multiple disorder realizations and ensemble averaging, we quantify the robustness of biphoton entanglement under realistic fabrication conditions.

\paragraph{Numerical solver and validation.}
We numerically integrate Eq.~\eqref{biphoton_schrodinger} using a symmetrized split-step exponential scheme with adaptive step-size control, maintaining a global integration error below \(10^{-6}\)~\cite{Agrawal2013,Press2007}. Validation was carried out by cross-checking all results against an independent second-order finite-difference Crank–Nicolson implementation~\cite{Crank1947}, yielding agreement within \(0.2\%\) throughout the propagation length. All presented propagation profiles, parity-resolved biphoton populations, and disorder robustness analyses derive from this rigorously validated numerical framework, thus reliably connecting structural symmetry, pump conditions, and entanglement properties in our topological photonic superlattices~\cite{Christodoulides2003,ZakeriCLEO2024}.

\section*{Supplementary text}

\textbf{\underline{Supplementary Note 1: Topological Characterization of Superlattices: Winding Number Analysis}}

One-dimensional chiral-symmetric superlattices generalize the celebrated Su–Schrieffer–Heeger (SSH) model~\cite{Su1979} by incorporating multiple waveguide sites within each unit cell. Increasing the number of sites per unit cell enriches the energy spectrum, introducing additional distinct bands, and enabling precise control of spectral bandgaps. By carefully adjusting intercell and intracell coupling strengths, we systematically tailor the lattice band structure, thereby modulating the transmission spectrum and achieving precise control over the operating bandwidth. Such meticulous spectral engineering holds promise for advanced photonic components, including disorder-robust waveguides and integrated entangled-photon sources.

Chiral symmetric superlattices inherently support symmetry-protected topological phases characterized by robust boundary states. Due to chiral symmetry, their energy spectra are symmetric about zero, enabling the topological phase to be quantified by an integer winding number \(\nu\)~\cite{Delplace2011,Asboth2016}. Physically, \(\nu\) enumerates the number of times the bulk Hamiltonian’s eigenstates encircle the origin in the complex plane as the Bloch wavevector \(k\) traverses the Brillouin zone. A nonzero winding number \(\nu \neq 0\) ensures the presence of topologically protected edge modes localized at lattice boundaries, resistant to symmetry-preserving perturbations and disorder~\cite{Atala2013}.

We analytically derive explicit expressions for the winding number and identify the conditions for gap closure in superlattices comprising three to six sites per unit cell (\(J=3,4,5,6\))~\cite{PhysRevA.98.043838}. These findings provide clear theoretical guidelines for selectively switching between topological and trivial phases, significantly broadening the functional capabilities of photonic lattice systems.

Through a suitable local permutation that groups all “\(\mathcal{A}\)” sites first, followed by all “\(\mathcal{B}\)” sites, the Bloch Hamiltonian for a \(J\)-site cell can be expressed in a block-off-diagonal form:
\[
H(k)=
\begin{bmatrix}
0_{n_A\times n_A} & q(k)\\[6pt]
q^\dagger(k) & 0_{n_B\times n_B}
\end{bmatrix},\qquad
\Gamma=\mathrm{diag}\bigl(\underbrace{+1,\dots,+1}_{n_A},\underbrace{-1,\dots,-1}_{n_B}\bigr),\quad
\{\Gamma,H\}=0.
\label{eq:chiral}
\]

The chiral symmetry places the system in class AIII of the Altland-Zirnbauer classification, for which the topological invariant in one dimension is the winding number:
\[
\nu=\frac{1}{2\pi}\int_{-\pi/a}^{\pi/a}\frac{\mathrm d}{\mathrm d k}\,\arg\det q(k)\;\mathrm d k.
\label{eq:windingNumber}
\]

For nearest-neighbor couplings, the elements of \(q(k)\) include only the first harmonic terms \(e^{\pm i k a}\), leading to a single-harmonic form of \(\det q(k)\):
\[
\det q(k)=C+R\,e^{-i k a},
\label{eq:singleHarmonic}
\]
with real constant \(C\) and positive constant \(R\) determined solely by coupling strengths. As the wavevector \(k\) sweeps through the Brillouin zone, \(\det q(k)\) traces a circle in the complex plane with radius \(R\) and center \(C\). Consequently, the winding number takes a simple, intuitive form:
\begin{equation}
\nu=
\begin{cases}
1, & \text{if}\quad R>|C|,\\[4pt]
0, & \text{if}\quad R<|C|.
\end{cases}
\label{eq:circleCriterion}
\end{equation}

The topological phase transition occurs at the critical condition \(R=|C|\), where the traced circle touches the origin, signaling closure of the central bulk gap (typically at wavevectors \(k=0\) or \(k=\pi/a\)). At this point, the winding number \(\nu\) can change, marking the boundary between distinct topological phases.

\textbf{\underline{Supplementary note 2: Topological bandgaps engineering and Edge‑mode localization}}

\begin{figure}[H]
\centering
\includegraphics[width=0.9\textwidth]{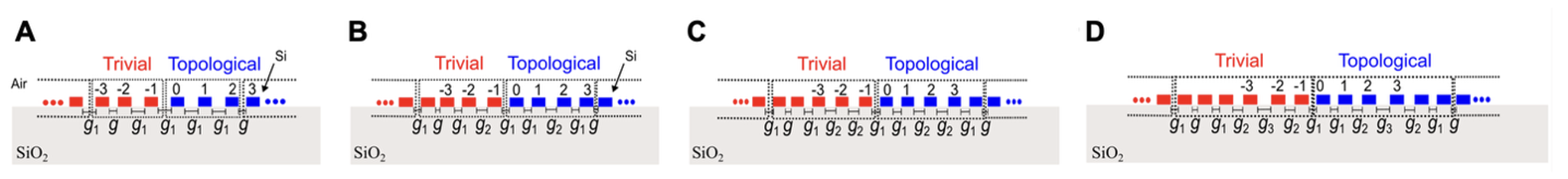}
\caption{Schematic of 1D photonic superlattice waveguide arrays. Panels (A–D) show lattices with \(M\) unit cells and \(J=3,4,5,6\) sites per cell, respectively. A central interface defines a defect region that governs the spatial localization of topological modes. The structure preserves chiral (sublattice) symmetry; each site on the right has a mirror partner on the left at the same distance from the center. A tight\hyp{}binding model with only nearest\hyp{}neighbor coupling is assumed, with intracell couplings \(t_{1},\ldots,t_{J-1}\) and intercell coupling \(\tau\).}
\label{structure}
\end{figure}

Consider a 1D periodic and discrete system which contain M unit cell and each unit cell compose J elements which illustrated in Fig.~\ref{structure}(A)-(D) for J=3, J=4, J=5 and J=6 respectively. The middle part of the array of the waveguides is referred to as the defect area. The defect area exerts a profound impact on the localization of the topological modes of the superlattices. Due to the chiral symmetry of this structure, each waveguide on the right side of the center has a corresponding waveguide on the left side at the same distance from the center. Using the tight-binding approximation, only the nearest-neighbor coupling is considered.

The corresponding Hamiltonian is:
\begin{equation}
H = \sum_{m=1}^{M} \sum_{j=1}^{J-1} t_j a_{m,j}^{\dagger} a_{m,j+1} + \tau a_{m,j}^{\dagger} a_{m+1,1} + h.c.,
\label{eq:Hamiltonian}
\end{equation}
where $a_{m,j}^{\dagger}$ ($a_{m,j}$) are is the creation (annihilation) operator at the $j$th site in the $m$th unit cell, $t_j$ is the intracell coupling amplitude from site $j$ to adjacent site $j + 1$, and $\tau$ is the intercell coupling amplitude (coupling amplitude between the unite cells) \cite{PhysRevA.98.043838}.

For the $J=3$ case, the Hamiltonian matrix in $k$-space is given by:
\[
H(k) = \begin{bmatrix}
0 & t_1 & \tau e^{-ika} \\
t_1 & 0 & t_2 \\
\tau e^{ika} & t_2 & 0
\end{bmatrix}
\quad k \in \left[-\frac{\pi}{a}, \frac{\pi}{a}\right].
\label{eq:HamiltonianMatrixJ3}
\]

Notice that $t_j = t_{J-j}=t$, due to the inversion symmetry. The dispersion relation is given by

\[
f(\beta) = \cos(ka) = \frac{(2t^2 + \tau^2)\beta - \beta^3}{2t^2 \tau},
\label{eq:DispersionRelationJ3}
\]
where $\beta$ represents the energy, $k$ the momentum, and $a$ the lattice constsnt. We can derive $\beta(k)$ from the above formula as
\[
\beta(k) = \left\{ -\frac{2^{1/3} B}{3A} + \frac{A}{3 \cdot 2^{1/3}},\ \frac{(1+i\sqrt{3})B}{3 \cdot 2^{2/3} A} - \frac{(1-i\sqrt{3})A}{6 \cdot 2^{1/3}},\ \frac{(1-i\sqrt{3})B}{3 \cdot 2^{2/3} A} - \frac{(1+i\sqrt{3})A}{6 \cdot 2^{1/3}} \right\},
\label{eq:BetaK}
\]
where $A$ and $B$ are defined as follows:
\[
A = \left(-54 \cos(ka) t^2 \tau + \sqrt{2916 t^4 \tau^2 \cos^2(ka) + 4B^3}\right)^{1/3}, \quad
B = -6t^2 - 3\tau^2.
\]

This yields three energy bands with two bandgaps. We obtain the band edges by solving $|f(\beta)|=1$ as
\[
|f(\beta)| = 1 \rightarrow \beta_1^{\pm} = \pm \frac{1}{2} (\tau + T), \quad \beta_2^{\pm} = \pm \tau, \quad \beta_3^{\pm} = \pm \frac{1}{2} (-\tau + T),
\]
where $T = \sqrt{8t^2 + \tau^2}$. The energy gaps $\Delta\beta = |\beta_j^+ - \beta_{j+1}^-|$ between the $j$th and $(j+1)$th bands are given by
\[
\Delta\beta_{12} = \Delta\beta_{23} = \frac{3}{2} \tau - \frac{1}{2} T = \frac{1}{2}(3\tau - \sqrt{8t^2 + \tau}).
\]

Due to symmetry, the gaps remain identical and close when $\tau = t$. This observation aligns with expectations, as equal intercell and intracell coupling simply leads to diffraction phenomena.

In the \( J=3 \) superlattice, the Bloch Hamiltonian takes the form of a \( 3 \times 3 \) Hermitian matrix.  Leveraging the lattice symmetries, however, it can be recast into a compact \(2 \times 2\) pseudospin form expanded on the Pauli matrices. This is achieved by choosing a suitable 2-dimensional representation—often through symmetry arguments or block diagonalization—that reduces the original \(3 \times 3\) Hamiltonian to an effective \(2 \times 2\) pseudospin Hamiltonian, capturing the essential physics in a simplified spinor space.

By defining $z = e^{ika}$ and $r = \frac{\tau}{t}$, the Hamiltonian simplifies to:
\[
H(k) = t
\begin{bmatrix}
0 & 1 & r z^{-1} \\
1 & 0 & 1 \\
r z & 1 & 0
\end{bmatrix}.
\]

This Hamiltonian can be represented as an effective two-level system using Pauli matrices $\sigma_x$ and $\sigma_y$:
\[
H(k) = t\left[ \mathrm{Re}(f(k)) \sigma_x + \mathrm{Im}(f(k)) \sigma_y \right], \quad f(k) = 1 + z + r z^2,
\]
here, $f(k)$ encodes momentum-dependent hopping amplitudes and phases, yielding a direct energy gap:
\[
\Delta(k) = 2t |f(k)|.
\]

The gap closes when $f(k)=0$, leading to the condition:
\[
1 + z + r z^2 = 0, \quad |z|=1.
\]

Separating real and imaginary parts yields critical topological transition points:
\begin{itemize}
    \item For $k=0$: $r = -2$ ($\tau = -2t$).
    \item For $k=\pi$: $r = 0$ ($\tau = 0$).
    \item For $r=1$: gap closes at $k = \pm \frac{2\pi}{3}$ ($\tau = t$).
\end{itemize}

Away from these special points, solutions for $z$ are:
\[
z_{\pm} = \frac{-1 \pm \sqrt{1 - 4r}}{2r}.
\]

The winding number $\nu(r)$, indicating the topological phase, equals the count of these solutions inside the unit circle:
\[
\nu(r) = \#\{|z_{\pm}(r)| < 1\}.
\]

Therefore,
\[
v=\begin{cases}
1, & |\tau|>|t|,\\[2pt]
0, & |\tau|<|t|.
\end{cases}
\]

Crossing any of the critical points (\( |\tau|=|t| \)) pushes a root through the unit circle, changing the winding and closing the bulk gap exactly at the corresponding momenta.

\begin{figure}[H]
\centering
\includegraphics[width=0.9\textwidth]{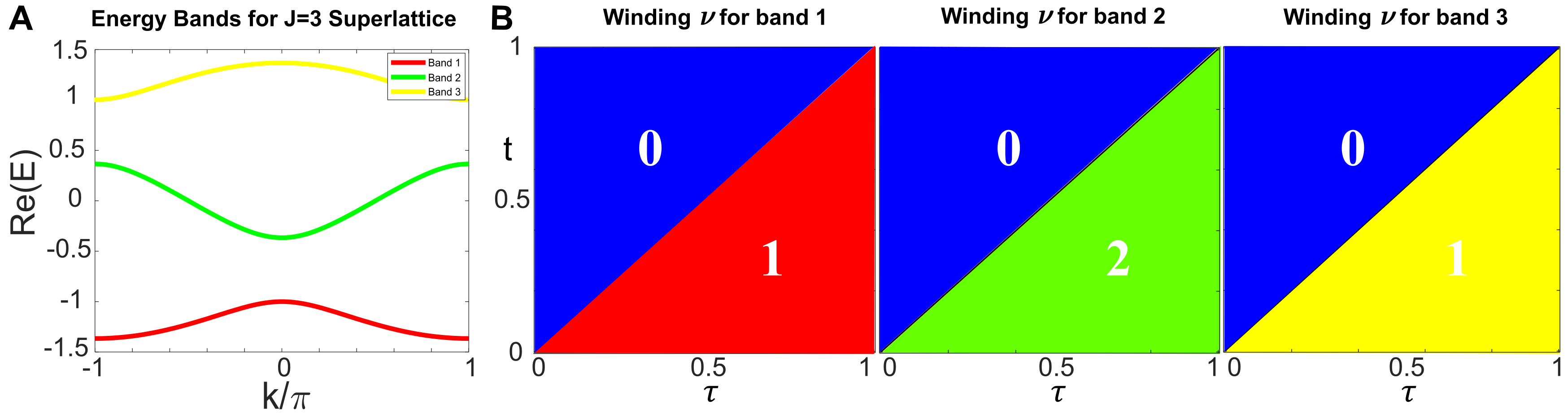}
\caption{Band structure and topological invariants of the superlattice $J=3$. (A) Real parts of the Bloch eigenvalues $E_n(k)$ across the first Brillouin zone ($k/\pi\in[-1,1]$) for representative couplings $t=0.5$ and $\tau=1$. (B) Band-resolved topological phase diagrams in the $(\tau,t)$ plane, obtained from numerically evaluated Berry phases.
Blue regions denote trivial topology ($\nu_n=0$); colored regions denote non-trivial phases (labels indicate $\nu_n$).
The sole phase boundary is the bulk gap-closure line $t=\tau$: for $\tau>t$ the three bands carry $(\nu_1,\nu_2,\nu_3)=(1,2,1)$, whereas for $\tau<t$ all three windings vanish.}
\label{winding}
\end{figure}

By sweeping the intracell and intercell couplings \(t,\tau\in[0,1]\), we obtain the band-resolved winding numbers of the superlattice \(J=3\), as shown in Fig.~\ref{winding}. Panel~A displays a representative Bloch dispersion at \(t=0.5\) and \(\tau=1\), deep in the topological regime \(\tau>t\). Panel~B charts the winding numbers \(\nu_n\) across the \((\tau,t)\) plane, evaluated from the Berry phase accumulated by the Bloch eigenvectors around the Brillouin zone. For \(\tau>t\) the spectrum is topological with \((\nu_1,\nu_2,\nu_3)=(1,2,1)\): the lower and upper bands carry unit winding, while the middle band winds twice about its singularities. For \(\tau<t\) all windings vanish, \((0,0,0)\), indicating a trivial phase. Topological transitions occur exclusively at bulk gap closures along the diagonal \(\tau=t\), which delineates the boundary between trivial and nontrivial regions.

The same root-tracking argument that gives the winding number also yields a closed-form decay length for the edge state. Writing the Bloch momentum at the band edge as \( k = \pi + i\kappa \) pushes one root \( z_\mathrm{in} \) of \( f(k) = 0 \) inside the unit circle to \( z_\mathrm{in} = e^{-\kappa} \); the evanescent envelope therefore falls off as \( \exp(-na/\xi) \), with
\[
    \xi(r) = \frac{a}{\left| \ln \left| z_\mathrm{in}(r) \right| \right|} = \frac{a}{\left| \ln \left| \frac{\tau}{t} \right| \right|}, \qquad r = \frac{t}{\tau},\quad t > 0.
\]

The decay length \( \xi \) quantifies how rapidly an edge mode fades into the bulk: the field amplitude drops as \( \exp(-n a / \xi) \) after \( n \) unit cells. A small \( \xi \) indicates that the mode is tightly confined, which enables compact photonic chips, efficient coupling to access waveguides, and enhanced immunity to fabrication disorder. Conversely, a large \( \xi \) implies weaker confinement, necessitating larger devices and making the mode more sensitive to imperfections. Thus, \( \xi \) serves as a key design parameter for optimizing device footprint, coupling efficiency, and topological robustness in photonic lattices.

The Zak phase for a 1D Bloch band \(n\) is the Berry phase over the Brillouin zone,
\[
\gamma_n=\oint_{\mathrm{BZ}} i\,\langle u_{n,k}\mid\partial_k u_{n,k}\rangle\,dk
\quad (\mathrm{mod}\ 2\pi).
\]

With inversion symmetry, \(\mathcal P H(k)\mathcal P^{-1}=H(-k)\), \(\gamma_n\) is quantized to \(0\) or \(\pi\) and can be read off at the time-reversal-invariant momenta (TRIMs) \(k=0,\pi/a\) using the parity eigenvalues \(\xi_n(k_i)=\pm1\):
\[
\gamma_n=\arg\!\big[\xi_n(0)\,\xi_n(\pi/a)\big]\quad (\mathrm{mod}\ 2\pi).
\]

For the \(3\times3\) superlattice
\[
\qquad
\mathcal P=
\begin{pmatrix}
0&0&1\\
0&1&0\\
1&0&0
\end{pmatrix},
\]
inversion swaps sites \(1\!\leftrightarrow\!3\) and fixes site \(2\). The TRIMs are the momenta satisfying \(k\equiv -k\) up to a reciprocal vector, i.e., \(k=0,\pi/a\). At these points \([\mathcal P,H(k_i)]=0\), so eigenstates can be chosen with definite parity.

Introduce the parity basis \(|S\rangle=(|1\rangle+|3\rangle)/\sqrt2\) and \(|M\rangle=|2\rangle\) (even), \(|A\rangle=(|1\rangle-|3\rangle)/\sqrt2\) (odd). With the unitary
\[
T=\big[\,|S\rangle,\ |M\rangle,\ |A\rangle\,\big],
\]
the TRIM Hamiltonians
\[
H(0)=
\begin{pmatrix}
0&t&\tau\\
t&0&t\\
\tau&t&0
\end{pmatrix},
\qquad
H(\pi/a)=
\begin{pmatrix}
0&t&-\tau\\
t&0&t\\
-\tau&t&0
\end{pmatrix}.
\]

Block diagonalize as
\[
T^\dagger H(0) T=
\begin{pmatrix}
\ \ \tau & \sqrt2\,t & 0\\
\sqrt2\,t & 0 & 0\\
0 & 0 & -\tau
\end{pmatrix},
\qquad
T^\dagger H(\pi/a) T=
\begin{pmatrix}
-\tau & \sqrt2\,t & 0\\
\sqrt2\,t & 0 & 0\\
0 & 0 & \ \ \tau
\end{pmatrix}.
\]

Thus the odd-parity eigenvalue is \(-\tau\) at \(k=0\) and \(+\tau\) at \(k=\pi/a\); the even-parity pair comes from the \(2\times2\) block with eigenvalues
\[
E_{\mathrm{even}}^{\pm}(0)=\frac{\tau\pm\sqrt{\tau^2+8t^2}}{2},
\qquad
E_{\mathrm{even}}^{\pm}(\pi/a)=\frac{-\tau\pm\sqrt{\tau^2+8t^2}}{2}.
\]

Comparison of the order of levels at \(k=0\) and \(k=\pi/a\) shows a single band inversion when \(|\tau|/|t|=1\). Reading the parity labels at the two TRIMs and inserting them in the TRIM formula for \(\gamma_n\) yields
\[
\gamma_{\text{lowest}}=\gamma_{\text{top}}=
\begin{cases}
0,&|\tau|<|t|,\\[2pt]
\pi,&|\tau|>|t|,
\end{cases}
\qquad
\gamma_{\text{middle}}=0\quad (t,\tau\neq0).
\]

which matches the bulk transition at \(|\tau|=|t|\) and predicts edge states only in the \(|\tau|>|t|\) phase. The Zak phase is gauge-defined modulo \(2\pi\), but its \(0/\pi\) quantization and the associated polarization \(P=\Gamma/(2\pi)\in\{0,\tfrac12\}\) are fixed by inversion symmetry as long as the gap remains open.

The same procedure was applied for $J = 4$, with the Bloch Hamiltonian given by
\[
H(k) = 
\begin{bmatrix}
0 & t_1 & 0 & \tau e^{-ika} \\
t_1 & 0 & t_2 & 0 \\
0 & t_2 & 0 & t_1 \\
\tau e^{ika} & 0 & t_1 & 0 \\
\end{bmatrix}.
\]

The dispersion relation is
\[
f(\beta) = \cos(ka) = \frac{\beta^4 - (2t_1^2 + t_2^2 + \tau^2)\beta^2 + t_1^4 + t_2^2 \tau^2}{2t_1^2 t_2 \tau}.
\]

The band edges are obtained from $|f(\beta)| = 1$:
\[
|f(\beta)|=1 \implies 
\begin{cases}
\beta_1^{\pm} = \frac{\pm t_2 - (T + \tau)}{2}, \\
\beta_2^{\pm} = \frac{t_2 \pm (T - \tau)}{2}, \\
\beta_3^{\pm} = -\beta_2^{\mp}, \\
\beta_4^{\pm} = -\beta_1^{\mp},
\end{cases}
\]
where $T = \sqrt{4t_1^2 + (t_2 - \tau)^2 }$. The energy gaps $\Delta\beta = |\beta_j^+ - \beta_{j+1}^-|$ are given by
\[
\Delta\beta = |\beta_j^+ - \beta_{j+1}^-| \implies 
\begin{cases}
\Delta\beta_{12} = \Delta\beta_{34} = |t_2 - \tau|, \\
\Delta\beta_{23} = |t_2 + \tau - T|.
\end{cases}
\]

\begin{figure}[H]
    \centering
    \includegraphics[width=0.9\textwidth]{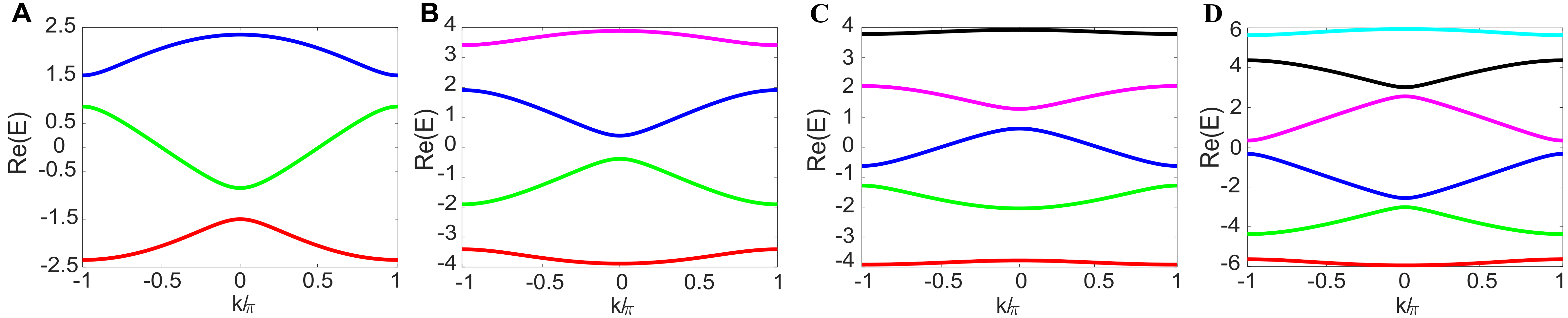}
    \caption{Energy band gap diagrams of the superlattice for fixed \(t\) and \(\tau\):
    (A) \(J=3\), (B) \(J=4\), (C) \(J=5\), and (D) \(J=6\).}
    \label{band}
\end{figure}

This indicates that the closure of the first and third band gaps occurs at $\tau=t_2$, whereas the second gap closes at $\tau=\frac{t_1^2}{t_2}$. Consequently, there are critical constraints on the intercell coupling parameter ($\tau>\max(t_2, \frac{t_1^2}{t_2})$) to have topological modes that must be considered when designing the superlattice structure. (the energy band gap diagram for different J is shown in Fig. \ref{band})

The decay length $\xi$ follows from analytically continuing the Bloch momentum to
$k_{\text{edge}}+i\kappa$, which pushes one root of $f(\beta)=0$ inside the unit
circle so that the edge--state envelope decays as $\exp(-n a/\xi)$.
Evaluating $f(\beta)=\pm1$ at the two independent band edges yields
\[
\xi_{(1)}=\frac{a}{\bigl|\ln\!\bigl|\tfrac{t_{2}}{\tau}\bigr|\bigr|},\qquad
\xi_{(0)}=\frac{a}{\bigl|\ln\!\bigl|\tfrac{t_{1}^{2}}{\tau t_{2}}\bigr|\bigr|},
\]
where $\xi_{(1)}$ diverges as $\tau\!\to\!t_{2}$ (closure of the outer gaps)
and $\xi_{(0)}$ diverges as $\tau\!\to\!t_{1}^{2}/t_{2}$ (closure of the
central gap).  Away from these critical points both decay lengths remain
finite, scaling inversely with their respective gap sizes and therefore
dictating the spatial confinement of the topological edge modes.

A five‑site superlattice with intracell hopping \(t_{1},t_{2}\) and an intercell hopping \(\tau\) is described in momentum space by
\[
H(k)=
\begin{bmatrix}
0 & t_{1} & 0 & 0 & \tau e^{-ika}\\
t_{1} & 0 & t_{2} & 0 & 0\\
0 & t_{2} & 0 & t_{2} & 0\\
0 & 0 & t_{2} & 0 & t_{1}\\
\tau e^{ika} & 0 & 0 & t_{1} & 0
\end{bmatrix}
t_{1},t_{2},\tau\in\mathbb{R}\setminus\{0\}.
\]

Chiral symmetry forces the spectrum to be symmetric about \(E=0\).  Expanding \(\det[H(k)-E]\) yields
\[
\det[H(k)-E]=E^{5}+c_{3}E^{3}+c_{1}E-c_{0}(k),
\]
with coefficients
\[
c_{3}=-(2t_{1}^{2}+2t_{2}^{2}+\tau^{2}),\qquad
c_{1}=t_{1}^{4}+2t_{1}^{2}t_{2}^{2}+2t_{2}^{2}\tau^{2},\qquad
c_{0}(k)=2t_{1}^{2}t_{2}^{2}\tau\cos(ka).
\]

Only the constant term depends on the crystal momentum, so setting \(E=0\) forces \(c_{0}(k)=0\), which in turn requires \(\cos(ka)=0\).  Consequently, regardless of the actual hopping amplitudes, the central band necessarily crosses zero energy at the high‑symmetry points \(k=\pm\pi/(2a)\); this zero mode is symmetry‑protected and cannot be removed without breaking chiral symmetry. To test whether any other two bands can touch, factor out the inevitable zero root:
\[
\det[H(k)-E]=E\bigl(E^{4}+\alpha E^{2}+\beta\bigr)-c_{0}(k),
\qquad
\alpha=-2(t_{1}^{2}+t_{2}^{2})-\tau^{2},\quad
\beta=t_{1}^{4}+2t_{1}^{2}t_{2}^{2}+2t_{2}^{2}\tau^{2}.
\]

The band touching some non-zero energy \(E_{\star}\) and the wave vector \(k_{\star}\) demands the simultaneous disappearance of the determinant and its energy derivative.  Imposing these conditions forces \(c_{0}(k_{\star})=0\) (so \(k_{\star}\) again satisfies \(\cos(ka)=0\)) and requires the quartic \(x^{2}+\alpha x+\beta=0\) in \(x=E^{2}\) to possess a repeated root.  The latter criterion is equivalent to the vanishing of the discriminant
\[
\Delta=\alpha^{2}-4\beta=\tau^{4}+4\tau^{2}\bigl(t_{1}^{2}-t_{2}^{2}\bigr)+4t_{2}^{4}.
\]

Introducing \(X=\tau^{2}\ge 0\) recasts \(\Delta=0\) as the quadratic equation
\[
X^{2}+4(t_{1}^{2}-t_{2}^{2})X+4t_{2}^{4}=0.
\]

Its own discriminant is \(16t_{1}^{2}(t_{1}^{2}-2t_{2}^{2})\).  If \(t_{1}^{2}<2t_{2}^{2}\) this discriminant is negative, so no real \(X\) exists; if \(t_{1}^{2}\ge 2t_{2}^{2}\) the quadratic does have roots, but a direct inequality check shows both are negative, contradicting the physical requirement \(X=\tau^{2}\ge 0\).  Hence there is no real solution for \(\Delta=0\) for any non‑zero real hoppings.


For a five‑site superlattice cell, chiral symmetry arranges
the Bloch Hamiltonian as
\(H_{5}(k)=\bigl(\begin{smallmatrix}0 & Q_{5}(k)\\ Q_{5}^{\dagger}(k) & 0
\end{smallmatrix}\bigr)\) where
\[
  Q_{5}(k)=
  \begin{bmatrix}
      t_{1} & 0   & \tau e^{-ik}\\
      t_{2} & t_{2} & 0          \\
      0     & t_{1} & t_{1}
  \end{bmatrix},\qquad
  q_{5}(k)=\det Q_{5}(k)=t_{1}^{2}t_{2}-\tau t_{2}^{2}e^{-ik}.
\]

The complex trajectory of \(q_{5}(k)\) is a circle centred at
\(t_{1}^{2}t_{2}-\tfrac12\tau t_{2}^{2}\) with radius
\(\tfrac12\tau t_{2}^{2}\); the origin lies inside this circle once
\(\tau>t_{1}^{2}/t_{2}\).  However, gap closure also requires inspection
of the time‑reversal invariant momentum \(k=\pi\), by imposing a condition such as \(\frac{t_1}{t_2} = \frac{4}{3}\), we can derive an explicit criterion for the bandgap closure. Expanding the secular
polynomial
\(P_{5}(\lambda)=\det(\lambda\openone-H_{5}(\pi))\) yields
\[
P_{5}(\lambda)=(-\lambda^{2}+\tau\lambda+t_{1}^{2})
               (-\lambda^{3}-\tau\lambda^{2}
                 +\lambda t_{1}^{2}+2\lambda t_{2}^{2}+2t_{2}^{2}\tau).
\]

A common root of the quadratic and cubic factors imposes
\[
  (\tau^{2}-t_{1}^{2}-t_{2}^{2})
  \Bigl(\tau^{2}-\frac{t_{1}^{2}t_{2}^{2}}{t_{1}^{2}+t_{2}^{2}}\Bigr)=0,
\]
giving two critical couplings
\[
  \tau_{(1)}=\frac{t_{1}t_{2}}{\sqrt{t_{1}^{2}+t_{2}^{2}}},
  \qquad
  \tau_{(2)}=\sqrt{t_{1}^{2}+t_{2}^{2}}.
\]

The localization lengths of the edge state follow from
\(k=\pi+i\kappa\) in \(q_{5}(k)=0\):
\[
  \kappa_{(1)}=\operatorname{arcosh}\!\bigl(\tfrac{\tau}{\tau_{(1)}}\bigr),
  \qquad
  \kappa_{(2)}=\operatorname{arcosh}\!\bigl(\tfrac{\tau_{(2)}}{\tau}\bigr),
  \qquad
  \xi_{(i)}=1/\kappa_{(i)}.
\]

For the six‑site cell (\(J=6\)) with intracell hoppings \(\{t_{1},t_{2},t_{3}\}\) and intercell \(\tau\), by considering the inversion symmetry, the hamiltonain matrix can be express as:
\[
H(k) =
\begin{bmatrix}
0 & t_1 & 0 & 0 & 0 & \tau e^{-ika} \\
t_1 & 0 & t_2 & 0 & 0 & 0 \\
0 & t_2 & 0 & t_3 & 0 & 0 \\
0 & 0 & t_3 & 0 & t_2 & 0 \\
0 & 0 & 0 & t_2 & 0 & t_1 \\
\tau e^{ika} & 0 & 0 & 0 & t_1 & 0
\end{bmatrix}.
\]

The chiral block is
\[
  Q_{6}(k)=
  \begin{bmatrix}
      t_{1} & 0     & \tau e^{-ik}\\
      t_{2} & t_{3} & 0           \\
      0     & t_{2} & t_{1}
  \end{bmatrix},\qquad
  q_{6}(k)=t_{1}^{2}t_{3}-\tau t_{2}^{2}e^{-ik}.
\]

Setting \(k=\pi\) inside \(q_{6}(k)=0\) closes the central band pair
(bands~3–4) and fixes a central‑gap coupling
\[
  \tau_{\mathrm C}=\frac{t_{1}^{2}t_{3}}{t_{2}^{2}}.
\]

To locate additional closures we analyses the Hermitian product
\(M_{6}(k)=Q_{6}^{\dagger}(k)Q_{6}(k)\).  Its single‑line spectral
invariants are
\[
\operatorname{Tr} M_{6} = 2(t_{1}^{2} + t_{2}^{2}) + t_{3}^{2} + \tau^{2}, \quad
\det M_{6} = t_{1}^{4} t_{3}^{2} + t_{2}^{2} \tau^{2} (t_{1}^{2} + t_{3}^{2}) - 2 t_{1} t_{2}^{2} t_{3} \tau \cos k.
\]

The band edges coincide when the cubic discriminant is
\[
  \Delta=(\operatorname{Tr}M_{6})^{2}
          -4\operatorname{Tr}M_{6}^{2}
          +18\det M_{6},
\]
vanishes.  At \(k=\pi\) this discriminant factorizes neatly:
\[
  \Delta\bigl|_{k=\pi}=
  \bigl(t_{1}^{2}\tau-t_{3}\tau^{2}-t_{2}^{2}t_{3}+t_{3}^{2}\tau\bigr)^{2}
  \;\Xi,
  \qquad \Xi>0,
\]
so, the square term alone controls double‑root formation.  Setting it to
zero produces a quadratic in \(\tau\):
\[
  (t_{1}^{2}-t_{3}^{2})\,\tau^{2}
  +2t_{2}^{2}t_{3}\,\tau
  -(t_{1}^{2}-t_{3}^{2})\,t_{3}^{2}=0.
\]

Its positive solutions define inner and outer closures,
\[
  \tau_{\pm}=
  \frac{-(t_{1}^{2}-t_{3}^{2})
        \pm\sqrt{(t_{1}^{2}-t_{3}^{2})^{2}+4t_{2}^{2}t_{3}^{2}}}
       {2t_{3}},
  \qquad
  \tau_{-}<\tau_{+}.
\]

Collectively, the three critical couplings
\(\tau_{-}<\tau_{\mathrm C}<\tau_{+}\) slice parameter space into four
spectral regimes distinguished by whether zero, one, two, or three
topological band pairs have closed. Using the same analytic continuation applied for \( J=5 \), the decay length are given by
\[
  \kappa_{\mathrm C}=
  \ln\!\bigl(\tfrac{t_{1}^{2}t_{3}}{t_{2}^{2}\tau}\bigr),\qquad
  \kappa_{\pm}= \operatorname{arcosh}\!
                 \Bigl[\frac{\tau^{2}+t_{3}^{2}-t_{1}^{2}}
                             {2t_{2}t_{3}}\Bigr],\qquad
  \xi=a/\kappa.
\]

Closed–form expressions are derived for all five spectral gaps in the superlattice (\(J = 6\)).

\[
\det[\lambda I-M_{6}(k)]=\lambda^{3}+p\lambda^{2}+q\lambda+r=0,
\]
\[
\begin{aligned}
p &= -\!\bigl[\,2(t_{1}^{2}+t_{2}^{2})+t_{3}^{2}+\tau^{2}\bigr],\\[4pt]
q &= t_{1}^{4}+2t_{1}^{2}t_{2}^{2}+t_{2}^{4}
     +2t_{1}^{2}t_{3}^{2}+t_{3}^{2}\tau^{2}+2t_{2}^{2}\tau^{2},\\[4pt]
r &= -\!\bigl[t_{1}^{4}t_{3}^{2}+2t_{1}^{2}t_{2}^{2}t_{3}\tau
               +t_{2}^{4}\tau^{2}\bigr].
\end{aligned}
\]

Using Cardano’s method,
\[
Q=\frac{3q-p^{2}}{9},
\qquad
R=\frac{9pq-27r-2p^{3}}{54},
\qquad
\theta=\arccos\!\Bigl(\frac{R}{\sqrt{-Q^{3}}}\Bigr),
\]
the three eigenvalues of \(M\) are
\[
\lambda_{j}(k)=
-\frac{p}{3}-2\sqrt{-Q}\,
\cos\!\Bigl(\tfrac{\theta+2\pi j}{3}\Bigr),\qquad j=0,1,2.
\]

Because \(H(k)\) is chiral, its positive energies are
\(E_{j}(k)=\sqrt{\lambda_{j}(k)}\).

\[
\begin{aligned}
&\Delta_{1}(k) = 2E_{0}(k), \quad
\Delta_{2}(k) = 2\bigl[E_{1}(k)-E_{0}(k)\bigr], \quad
\Delta_{3}(k) = 2\bigl[E_{2}(k)-E_{1}(k)\bigr], \\
&\Delta_{4}(k) = \Delta_{3}(k), \quad
\Delta_{5}(k) = \Delta_{2}(k).
\end{aligned}
\]

At the end of this section, we present Table~\ref{tab:summary_superlattice}, which provides a compact summary of key analytical results for superlattice models with unit cell sizes \(J = 3, 4, 5, 6\), including gap-closure conditions, decay lengths, and gap size expressions. This concise representation serves as a practical reference, facilitating direct application of theoretical insights for designing robust photonic superlattices and enabling rapid identification of topological phases and critical transition points in these systems.

\begin{table}[htbp]
\centering
\renewcommand{\arraystretch}{1.5}
\caption{Summary of key properties for superlattices ($J=3,4,5,6$)}
\label{tab:summary_superlattice}
{\large
\resizebox{1\textwidth}{!}{%
\begin{tabular}{c|c|c|c}
\hline
\textbf{J} & \textbf{Gap Closure Conditions} & \textbf{Decay Length} & \textbf{Gap Size} \\
\hline\hline
3 & $\tau = t$ & $\xi=\frac{a}{|\ln|\frac{\tau}{t}||}$ & $\Delta\beta_{12}=\Delta\beta_{23}=\frac{1}{2}(3\tau-\sqrt{8t^2+\tau^2})$ \\[6pt] \hline
4 & \begin{tabular}[c]{@{}c@{}} $t_1^2 = t_2 \tau$ \\  $t_2 = \tau$\end{tabular} & 
\begin{tabular}[c]{@{}c@{}}$\xi_0=\frac{a}{|\ln|\frac{t_1^2}{\tau t_2}||}$\\[4pt] $\xi_1=\frac{a}{|\ln|\frac{t_2}{\tau}||}$\end{tabular} & 
\begin{tabular}[c]{@{}c@{}}$\Delta\beta_{12}=\Delta\beta_{34}=|t_2-\tau|$\\[4pt] $\Delta\beta_{23}=|t_2+\tau-\sqrt{4t_1^2+(t_2-\tau)^2}|$\end{tabular} \\[6pt] \hline
5 & \begin{tabular}[c]{@{}c@{}}$\tau_{(1)}=\frac{t_1 t_2}{\sqrt{t_1^2+t_2^2}}$\\[4pt] $\tau_{(2)}=\sqrt{t_1^2+t_2^2}$ \\[4pt] If $\frac{t_1}{t_2}=\frac{4}{3}$, above gap closure conditions are valid.\end{tabular} & 
\begin{tabular}[c]{@{}c@{}}$\xi_{(1)}=\frac{a}{\mathrm{arcosh}(\frac{\tau}{\tau_{(1)}})}$\\[4pt] $\xi_{(2)}=\frac{a}{\mathrm{arcosh}(\frac{\tau_{(2)}}{\tau})}$\end{tabular} & 
presented in the main text \\[6pt] \hline
6 & \begin{tabular}[c]{@{}c@{}} $\tau_{\mathrm{C}}=\frac{t_1^2 t_3}{t_2^2}$\\[4pt] $\tau_{\pm}=\frac{-(t_{1}^{2}-t_{3}^{2})\pm\sqrt{(t_{1}^{2}-t_{3}^{2})^{2}+4t_{2}^{2}t_{3}^{2}}}{2t_{3}}$\end{tabular} & 
\begin{tabular}[c]{@{}c@{}}$\xi_{\mathrm{C}}=\frac{a}{\ln(\frac{t_1^2 t_3}{t_2^2 \tau})}$\\[4pt] $\xi_{\pm}=\frac{a}{\mathrm{arcosh}\left[\frac{\tau^2+t_3^2-t_1^2}{2t_2 t_3}\right]}$\end{tabular} & 
presented in the main text\\[6pt] \hline
\end{tabular}}
}
\end{table}

\textbf{\underline{Supplementary note 3: Output power characterization}}

\begin{figure}[H]
\centering
\includegraphics[width=0.9\textwidth]{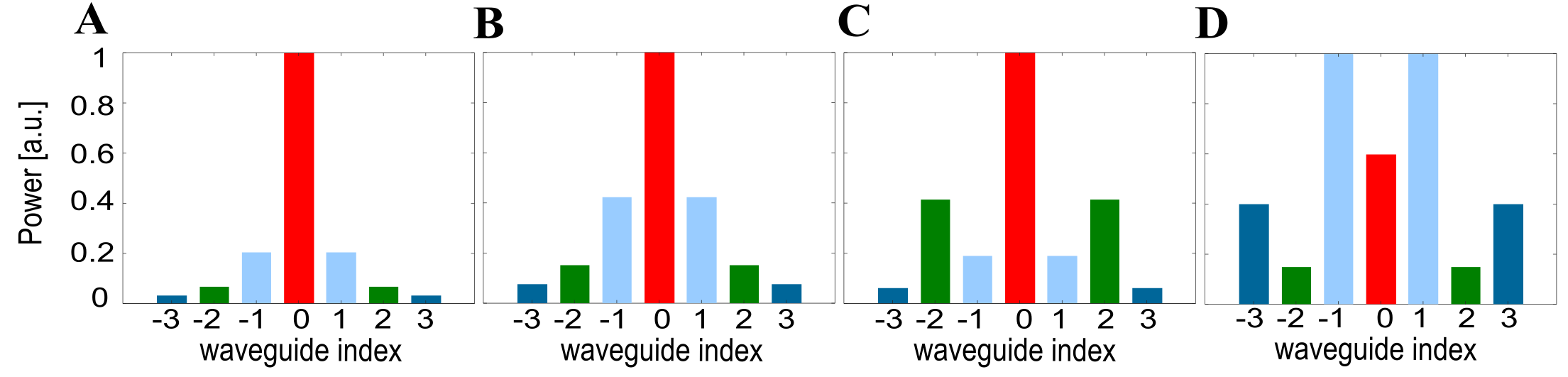}
\caption{Simulated on\hyp{}chip facet output power distributions across the seven read\hyp{}out waveguides for superlattices with \(J=3,4,5,6\) \textit{(A–D)}. Waveguides equidistant from the central channel are identically colored; the equal heights of color matched bars verify the mirror symmetry of the array. This behavior reflects the sublattice (chiral) symmetry of the tight binding Hamiltonian, which enforces equal modal amplitudes at sites mirrored about the interface.}
\label{Fig7}
\end{figure}

Figure~\ref{Fig7} depicts simulated output power profiles on the chip facet for seven read-out waveguides for superlattice \(J = 3-6\).  
The waveguides equidistant from the central channel share the same color, making symmetric pairs immediately recognizable.  
For every value of \(J\), the color-matched bars attain identical heights, confirming the mirror symmetry predicted by theory.  
This behavior reflects the sublattice (chiral) symmetry inherent to the tight-binding Hamiltonian, which enforces equal modal amplitudes at sites mirrored about the lattice interface.

\begin{figure}[H]
\centering
\includegraphics[width=0.9\textwidth]{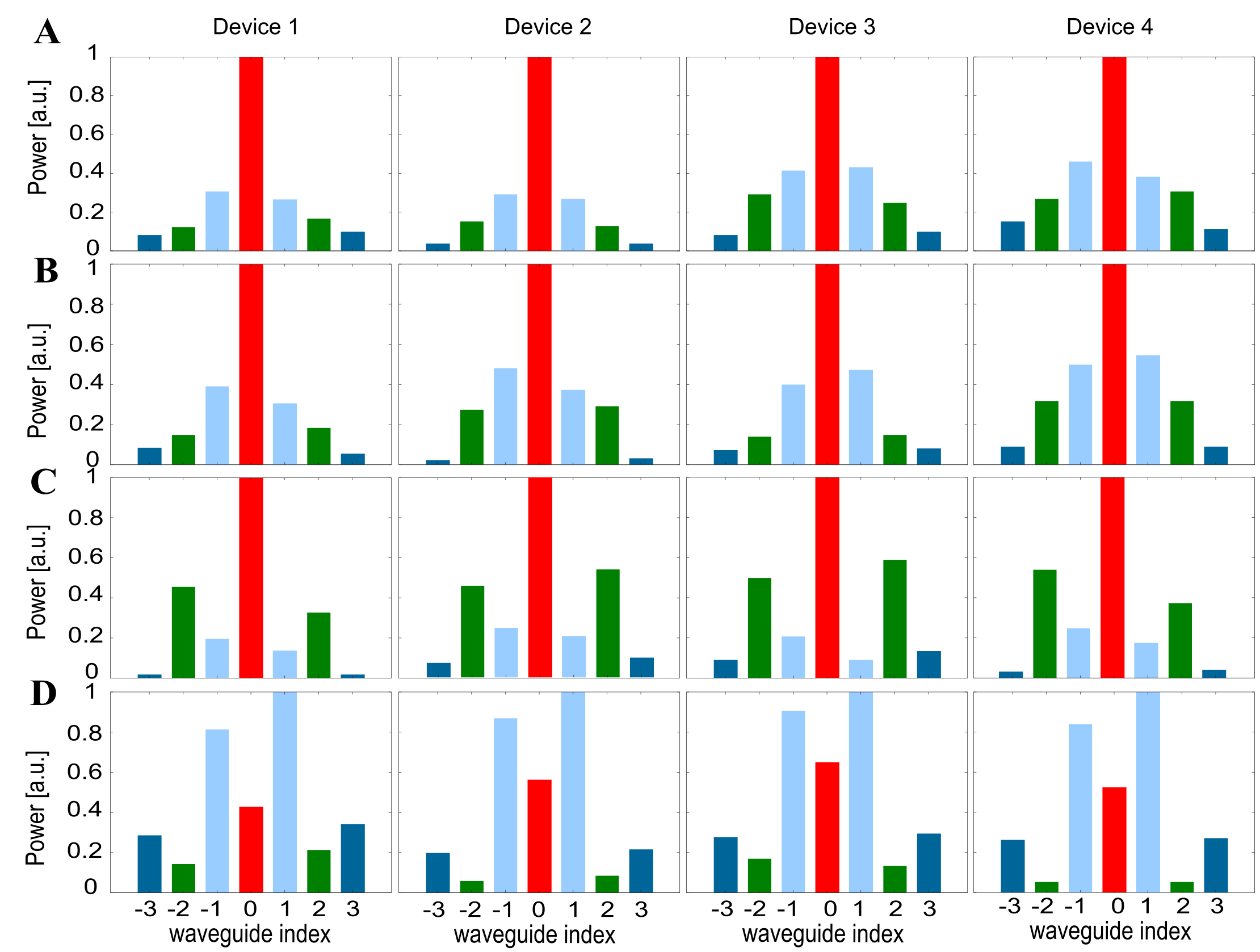}
\caption{Measured output-power distributions across seven waveguides for four independent devices fabricated by electron-beam lithography (A) J=3 (B) J=4 (C) J=5 (D) J=6. }
\label{Fig8}
\end{figure}

Experimental validation, presented in Fig.~\ref{Fig8}, demonstrates measured output power distributions from four nominally identical photonic devices (Devices 1--4) fabricated using high resolution electron beam lithography. Despite unavoidable minor variations in side-lobe intensities due to subtle differences in waveguide geometry, spacing, and refractive index profiles, the distinctive topological mode patterns remain consistently robust. Quantitative analysis reveals that the similarity between the experimental data and the ideal simulated profiles consistently exceeds 90\% in all devices tested. These results emphasize the inherent resilience of engineered topological interface modes to fabrication-induced imperfections, underscoring the significant potential of topological photonic superlattices for scalable, reliable generation, and precise manipulation of complex spatial modes in integrated photonic circuits.

\textbf{\underline{Supplementary Note 4: Biphoton Eigenmode Population}}

Expanding the biphoton state $\psi$ into the eigenmode basis yields:
\[
\psi(z) = \sum_{m,n} \psi_{mn}(z) \, |V_s(m)\rangle \langle V_i(n)|,
\]
where $V_s(m)$ and $V_i(n)$ are eigenvectors of the signal and idler Hamiltonians:
\[
H_s |V_s(m)\rangle = \beta_m^{(s)} |V_s(m)\rangle, \quad H_i |V_i(n)\rangle = \beta_n^{(i)} |V_i(n)\rangle.
\]

Projecting the operator equation onto a single eigen-pair $(m,n)$ by multiplying from the left by $\langle V_s(m)|$ and from the right by $|V_i(n)\rangle$ produces an ordinary differential equation:
\[
i \frac{d \psi_{mn}}{d z} = (\beta_m^{(s)} - \beta_n^{(i)}) \psi_{mn} + \gamma \psi_0 A_p^2(z) C_{mn},
\]
where the overlap constant is defined as:
\[
C_{mn}=\bigl[V_{\mathrm{s}}(m)\bigr]_{p}^{*}\,\bigl[V_{\mathrm{i}}(n)\bigr]_{p},
\]
with \(p\) denoting the pumped waveguide. In single–site pumping, \(C_{mn}\) depends on the mode profiles and can be absorbed into \(\psi_{0}\).
If the pump phase is retained inside the squared pump envelope,
\[
A_p^{2}(z)=A_{0}^{2}\,e^{i\,2\beta_{p}z}.
\]

Then the factor \(2\beta_{p}\) is absorbed into \(A_p^{2}\). Hence we may replace \(\beta_m^{(s)}-\beta_n^{(i)}-2\beta_p\) by \(\beta_m^{(s)}-\beta_n^{(i)}\).
Defining the mismatch \(\Delta\beta_{mn}=\beta_{m}^{(s)}-\beta_{n}^{(i)}\), the evolution equation simplifies to
\[
i\,\frac{d\psi_{mn}}{dz}
= \Delta\beta_{mn}\,\psi_{mn}
+ \gamma\,\psi_{0}\,A_p^{2}(z)\,C_{mn}.
\]

Employing the integrating factor \(e^{i\Delta\beta_{mn}z}\) with the initial condition \(\psi_{mn}(0)=0\), we obtain
\[
\psi_{mn}(L)=-i\,\gamma\,\psi_{0}\,C_{mn}\!\int_{0}^{L}\!A_p^{2}(z')\,
e^{-i\Delta\beta_{mn}(L-z')} \,dz'.
\]

For a constant, undepleted pump \(A_p^{2}(z)=A_p^{2}\), this integral evaluates analytically to:
\[
\psi_{mn}(L)=\frac{2\gamma\,\psi_{0}\,A_p^{2}\,C_{mn}}{\Delta\beta_{mn}}\;
e^{-i\Delta\beta_{mn}L/2}\,
\sin\!\left(\frac{\Delta\beta_{mn}L}{2}\right).
\]

Thus the biphoton eigenmode population is
\[
B_{mn}(L)=\bigl|\psi_{mn}(L)\bigr|^{2}
=\left|\frac{2\gamma\,\psi_{0}\,A_p^{2}\,C_{mn}}{\Delta\beta_{mn}}\,
\sin\!\left(\frac{\Delta\beta_{mn}L}{2}\right)\right|^{2},
\]
i.e., it follows a \(\bigl|\sin x/x\bigr|^{2}\) envelope.
For nearly phase–matched pairs \(\bigl|\Delta\beta_{mn}\,L\bigr|\ll 1\),
\[
B_{mn}(L)\approx \bigl[\gamma\,\psi_{0}\,A_p^{2}\,C_{mn}\,L\bigr]^{2}.
\]

For finite mismatch, the first zero and first maximum occur at
\[
L_{\text{zero}}=\frac{2\pi}{\lvert\Delta\beta_{mn}\rvert},\qquad
L_{\max}=\frac{\pi}{\lvert\Delta\beta_{mn}\rvert}.
\]

This formulation establishes the theoretical basis for numerical calculations of biphoton eigenmode populations, explicitly showing that the population dynamics are determined by both modal localization (setting initial nonlinear gain) and phase mismatch \(\Delta \beta_{mn}\) (governing long-range oscillatory behavior). Highly localized modes exhibit rapid initial growth, while mismatched modes display characteristic oscillations, periodically returning amplitude to the pump.

\textbf{\underline{Supplementary Note 5: Impact of Mode Localization on Entanglement}}

\begin{figure}[H]
\centering
\includegraphics[width=0.9\textwidth]{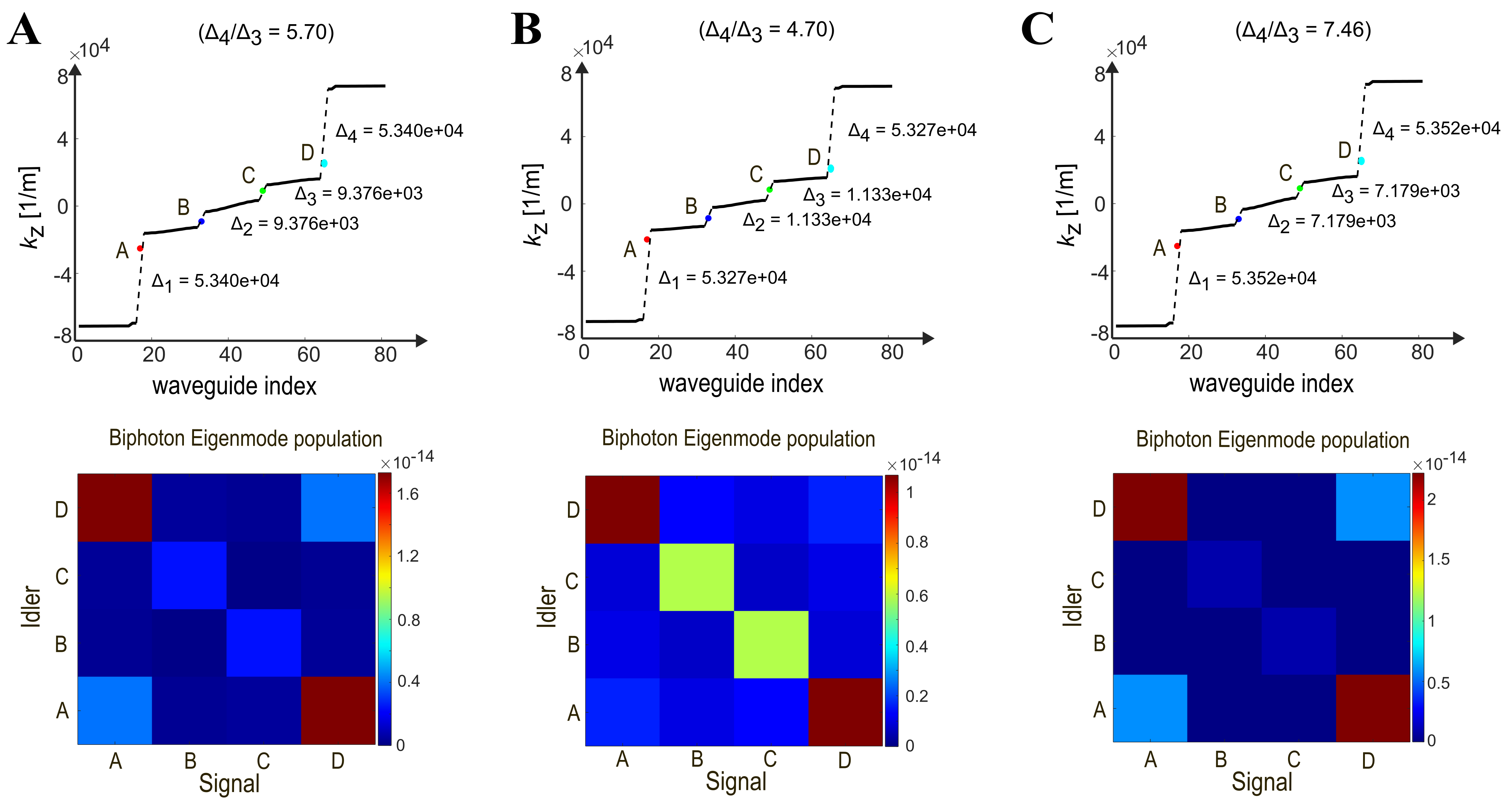}
\caption{Biphoton eigenmode population maps for a \(J=5\) superlattice after \(500~\mu\mathrm{m}\) propagation while sweeping the intracell gap \(t_{1}\) (with \(t_{2}=330~\mathrm{nm}\) and \(\tau=125~\mathrm{nm}\) fixed). Enlarging the central band gap \(\Delta_{3}\) strengthens localization of the central topological modes and increases their populations (modes B and C); narrowing \(\Delta_{3}\) has the opposite effect. (A) \(t_{1}=282~\mathrm{nm}\): \(\Delta_{3}=9.38\times10^{3}\), \(\Delta_{4}/\Delta_{3}=5.70\); (B) \(t_{1}=310~\mathrm{nm}\): \(\Delta_{3}=1.13\times10^{4}\), \(\Delta_{4}/\Delta_{3}=4.70\); (C) \(t_{1}=260~\mathrm{nm}\): \(\Delta_{3}=7.19\times10^{3}\), \(\Delta_{4}/\Delta_{3}=7.46\).}
\label{EigenMode}
\end{figure}

The distribution of biphoton eigenmode populations within the entangled state critically depends on mode localization, particularly for topological boundary modes whose localization is intimately related to the widths of their respective topological band-gaps. To elucidate this relationship, we systematically vary the central gap sizes in a $J = 5$ superlattice and analyze the resultant biphoton populations after a propagation distance of $500\,\mu\mathrm{m}$. Specifically, Fig.~\ref{EigenMode} demonstrates how adjustments in the intracell coupling parameter $t_1$, with other coupling parameters held constant, influence the size of the central gap $\Delta_3$ and consequently modify the gap ratio $\Delta_4 / \Delta_3$.

An increase in the central band-gap $\Delta_3$ enhances the localization of the corresponding central topological modes, thereby significantly improving their spatial overlap with the pump field. This intensified overlap leads to an increased population of these modes within the biphoton eigenmode basis (specifically modes B and C). Conversely, narrowing the central band-gap weakens the localization of these modes, diminishing their overlap with the pump and consequently reducing their contribution to biphoton generation.

In Fig.~\ref{EigenMode}(A), the original design parameters ($t_1 = 282\,\mathrm{nm}$, $t_2 = 330\,\mathrm{nm}$, $\tau = 125\,\mathrm{nm}$) yield a moderate central-gap width $\Delta_3 = 9.38 \times 10^3$, corresponding to a gap ratio $\Delta_4/\Delta_3 = 5.70$. This configuration results in a balanced occupation of the second and third topological modes. By increasing $t_1$ to $310\,\mathrm{nm}$, the central-gap size expands to $1.13 \times 10^4$, thus lowering the ratio $\Delta_4/\Delta_3$ to 4.70. This adjustment markedly elevates the eigenmode populations of modes 2 and 3, as evidenced by the pronounced intensities in the corresponding heat map entries. In contrast, reducing $t_1$ to $260\,\mathrm{nm}$ contracts the central gap $\Delta_3$ to $7.19 \times 10^3$, increasing the gap ratio to $\Delta_4/\Delta_3 = 7.46$, and noticeably suppressing the population of these modes.

\textbf{\underline{Supplementary note 6: Impact of modal phase-mismatch on entanglement}}

The initial growth of each biphoton–eigenmode population is primarily governed by the spatial confinement strength of the corresponding edge states within the topological bandgap. However, at longer propagation distances, the mode populations evolve according to the relative phase accumulated by the signal and idler photons, dictated by their modal phase mismatch.

For strongly mismatched mode pairs, the dependence of eigenmode‑population amplitude on propagation length~$L$ becomes sinusoidal rather than linear. Specifically, the population first peaks when
\[
\frac{\Delta\beta_{mn}L}{2} = \frac{\pi}{2},
\]
and then oscillates with spatial period
\[
L_{\text{osc}} = \frac{2\pi}{\lvert \Delta\beta_{mn}\rvert}\; .
\]

To illustrate this, we consider a representative $J = 6$ superlattice with coupling parameters chosen inside the topological regime. Numerical analysis yields the outermost edge‑mode phase mismatch
\[
\Delta\beta_{AE} \;=\; \beta_{A}^{(s)} - \beta_{E}^{(i)} \;\approx\; 2.6025\times10^{3}\,\mathrm{m^{-1}} .
\]

Because of this mismatch, the signal and idler photons accumulate opposing phases, producing complete destructive interference after the propagation length.
\[
L_{\text{zero}} 
\;=\; \frac{2\pi}{\lvert\Delta\beta_{AE}\rvert} 
\;\approx\; \frac{6.283}{2.6025\times10^{3}}
\;\approx\; 2.42\,\mathrm{mm}.
\]

\begin{figure}[H]
    \centering
    \includegraphics[width=0.8\textwidth]{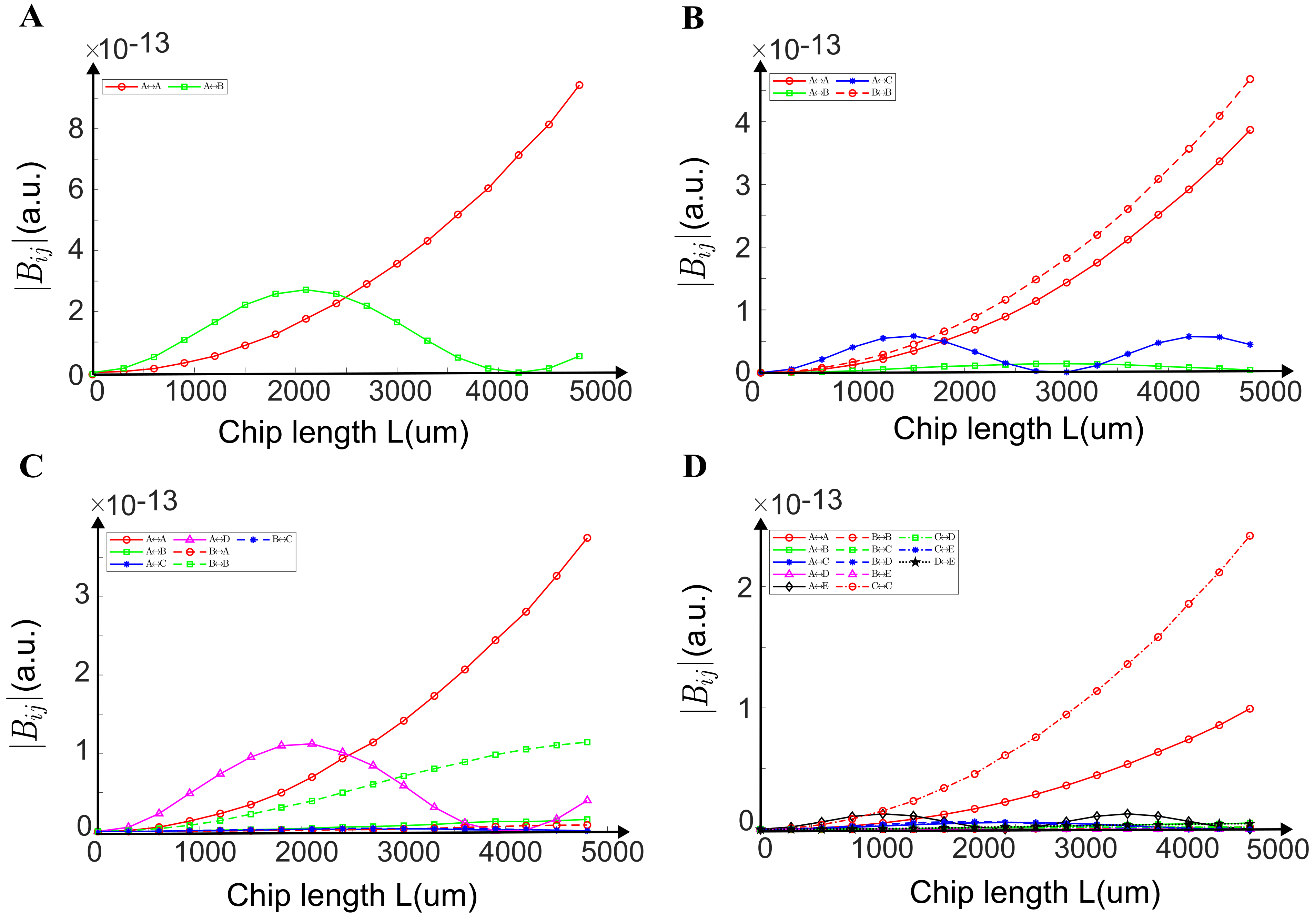} 
    \caption{Evolution of biphoton eigenmode populations versus propagation length \(L\) for superlattices with \(J=3,4,5,6\) \textit{(A–D)}. The initial rise reflects edge\hyp{}state localization, whereas the long\hyp{}range dynamics are governed by the modal phase mismatch \(\Delta\beta_{mn}\): negligible mismatch yields approximately quadratic growth \(\propto L^{2}\), while finite mismatch produces sinusoidal oscillations with \(L_{\mathrm{osc}}=2\pi/|\Delta\beta_{mn}|\) and a first maximum at \(\Delta\beta_{mn}L/2=\pi/2\). In (D), the outermost pair \(AE\) has \(\Delta\beta_{AE}\approx 2.6025\times10^{3}\,\mathrm{m^{-1}}\), implying \(L_{\mathrm{zero}}\approx 2.42~\mathrm{mm}\), consistent with the observed node near \(L\approx 2.47~\mathrm{mm}\).}
    \label{BEM_Propagation}
\end{figure}

This analytical prediction coincides with the drop in Fig.~\ref{BEM_Propagation}(D), where the $AE$ mode population reaches zero near $L \approx 2.47\,\mathrm{mm}$. Beyond this point, the amplitude is revived and continues to oscillate, revealing the characteristic phase‑mismatch dynamics.

Figure~\ref{BEM_Propagation} summarizes the biphoton eigenmode populations versus propagation length~$L$ for superlattices with $J=3$–$6$. In each case, the initial increase reflects spatial localization, while the long‑range behavior stems from the interplay between confinement and phase mismatch. Negligible mismatch leads to approximately quadratic growth ($\propto L^{2}$), whereas finite mismatch induces sinusoidal oscillations and periodic energy exchange with the pump. Hence, the balance between localization and mismatch determines whether a mode pair exhibits sustained amplification or persistent oscillatory dynamics. In particular, in even $J$ lattices ($J=4,6$) the larger mismatches cause earlier maxima and minima than in odd $J$ cases, while the near zero mismatch $CC$ mode grows monotonically and ultimately dominates for large~$L$.

\subsection*{\textbf{\underline{Supplementary Note 7: Impact of Modal Parity on Entanglement}}}

In a mirror-symmetric lattice, every Bloch eigenmode has definite parity symmetric $(+)$ or antisymmetric $(-)$ about the central interface. With the pump centered at the interface (even intensity profile), parity imposes a selection rule on spontaneous four-wave mixing (SFWM) and structures the biphoton entanglement (10). Let $f^{(s)}_{\alpha}(n)$ and $f^{(i)}_{\beta}(n)$ be the signal and idler eigenmode profiles with parities $\sigma_{\alpha},\sigma_{\beta}\in\{+1,-1\}$ (so that $f_{j}(-n)=\sigma_{j}f_{j}(n)$). In the undepleted-pump regime, the coupling amplitude into \((\alpha,\beta)\) is
\[
\eta_{\alpha\beta}\propto \sum_{n} A_{p}^{2}(n)\,f^{(s)}_{\alpha}(n)\,f^{(i)}_{\beta}(n).\
\]

with \(A_{p}^{2}(n)\) even in \(n\). Grouping terms by \(\pm n\) gives 
\[
\eta_{\alpha\beta}\propto \sum\limits_{n>0} A_{p}^{2}(n)\,\bigl[1+\sigma_{\alpha}\sigma_{\beta}\bigr]\,f^{(s)}_{\alpha}(n)\,f^{(i)}_{\beta}(n).
\]

So, mixed-parity pairs ($\sigma_{\alpha}\sigma_{\beta}=-1$) cancel identically, whereas same-parity pairs ($\sigma_{\alpha}\sigma_{\beta}=+1$) add constructively. A superlattice further refines this rule because antisymmetric modes can carry different phase-flip periods across the unit cell. For two antisymmetric modes, the product $f^{(s)}_{\alpha}f^{(i)}_{\beta}$ is even but may oscillate at the supercell scale; coupling is therefore strongest when the phase-flip periodicities match (or are commensurate) and is suppressed when they do not.

\begin{figure}[H]
    \centering
    \includegraphics[width=0.7\textwidth]{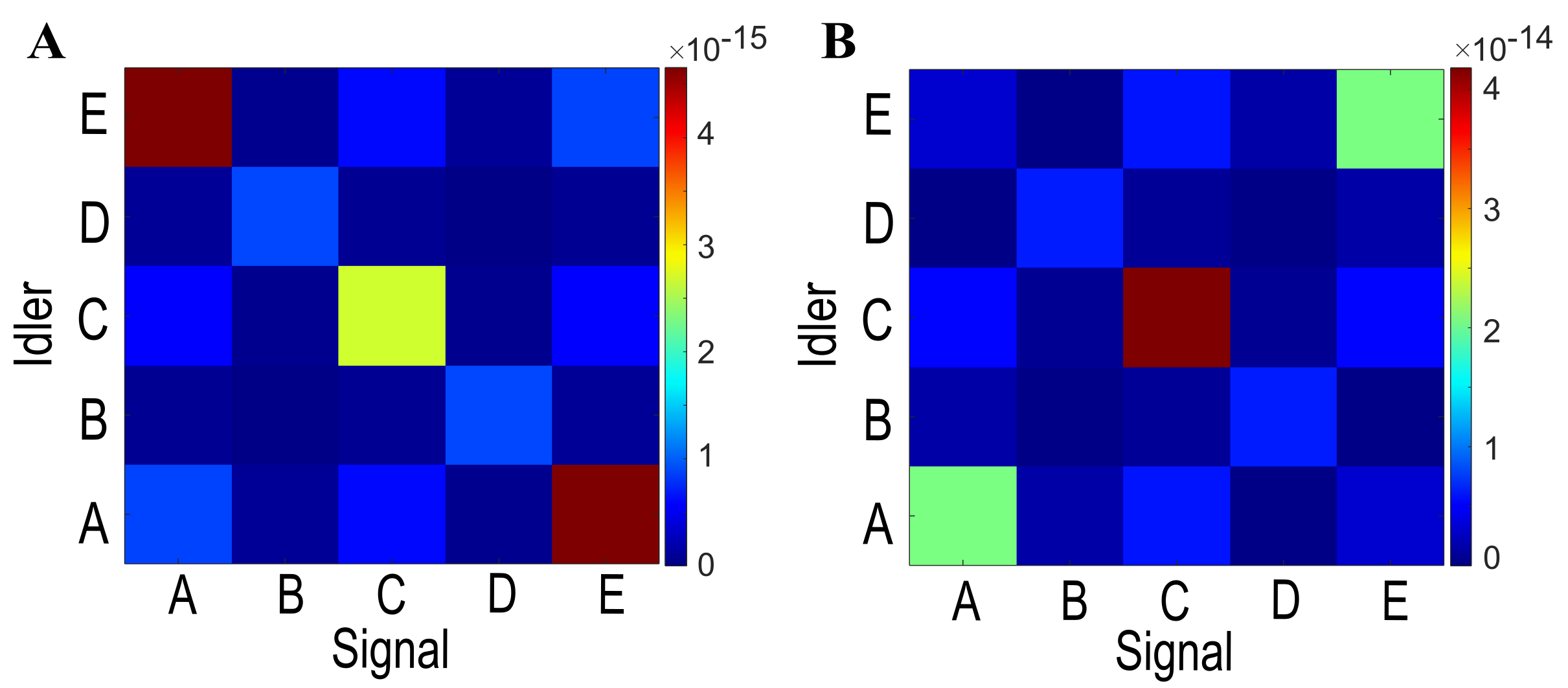}
     \caption{Parity\hyp{}filtered biphoton eigenmode populations in a mirror\hyp{}symmetric \(J=6\) superlattice with an interface\hyp{}centered (even) pump. Mixed\hyp{}parity pairs cancel, same\hyp{}parity pairs couple, and supercell phase\hyp{}flip periodicity further selects among antisymmetric pairs. (A) \(L=500~\mu\mathrm{m}\): parity\hyp{}allowed channels dominate; (B) \(L=2000~\mu\mathrm{m}\): modal beating redistributes weight while suppressed entries remain weak.}
    \label{parity}
\end{figure}

Figure \ref{parity} shows biphoton eigenmode populations for \(J=6\) at two propagation distances. At \(L=500~\mu\mathrm{m}\) [panel~A], population concentrates in parity-allowed channels with several entries near the background level, consistent with mixed-parity suppression. At \(L=2000~\mu\mathrm{m}\) [panel~B], modal beating redistributes weight among the allowed channels (e.g., an enhanced central \(CC\) element and symmetric corner pairs remain visible), while the many weak entries persist, again consistent with parity constraints and the superlattice periodicity filter. Together, these observations indicate that parity and supercell phase structure jointly govern which topological mode pairs are populated and how their weights evolve with \(z\), providing a symmetry-guided handle on the entanglement structure.

\textbf{\underline{Supplementary note 8: Impact of disorder on entanglement}}

\begin{figure}[H]
  \centering
  \includegraphics[width=1\textwidth]{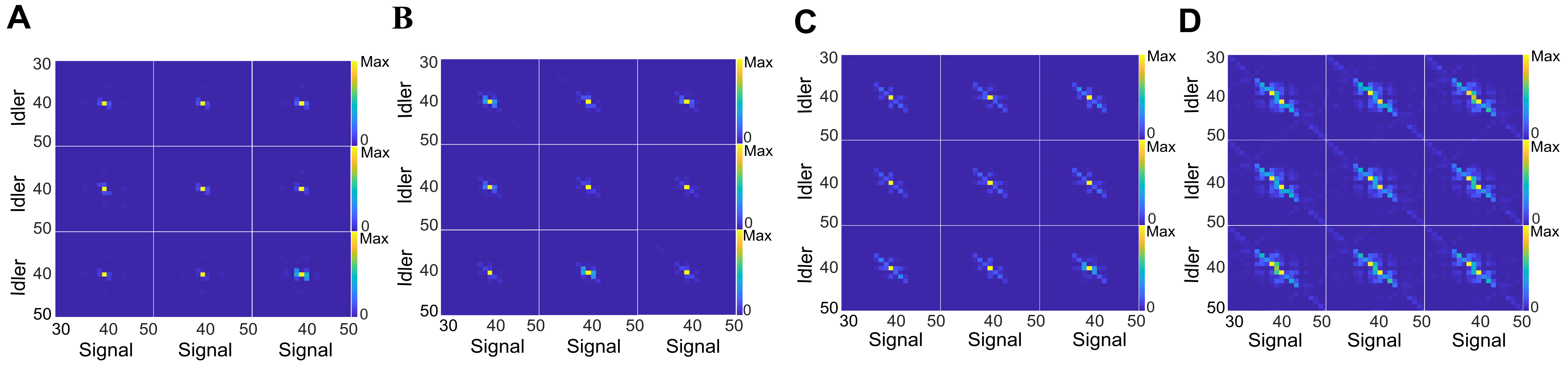}
  \caption{Simulated output biphoton correlations under disorder in the couplings. (A-D) Show nine independent disorder realizations, drawn from a 10\%-standard-deviation Gaussian distribution, for superlattice orders $J=3$–6, respectively.}
  \label{biphoton}
\end{figure}

To quantitatively assess the robustness of the entangled state against disorder in the couplings, we impose controlled off‑diagonal disorder by randomly perturbing the inter‑waveguide spacings, and perform propagation simulations. Coupling strengths were perturbed according to a Gaussian distribution with relative standard deviation of 10\%. Remarkably, the spatial correlation profiles shown in Fig.\ref{biphoton} (A-D) for the respective $J=3$–6 cases, remain essentially unchanged across nine random realizations of such disorder.

\begin{figure}[H]
\centering
\includegraphics[width=0.8\textwidth]{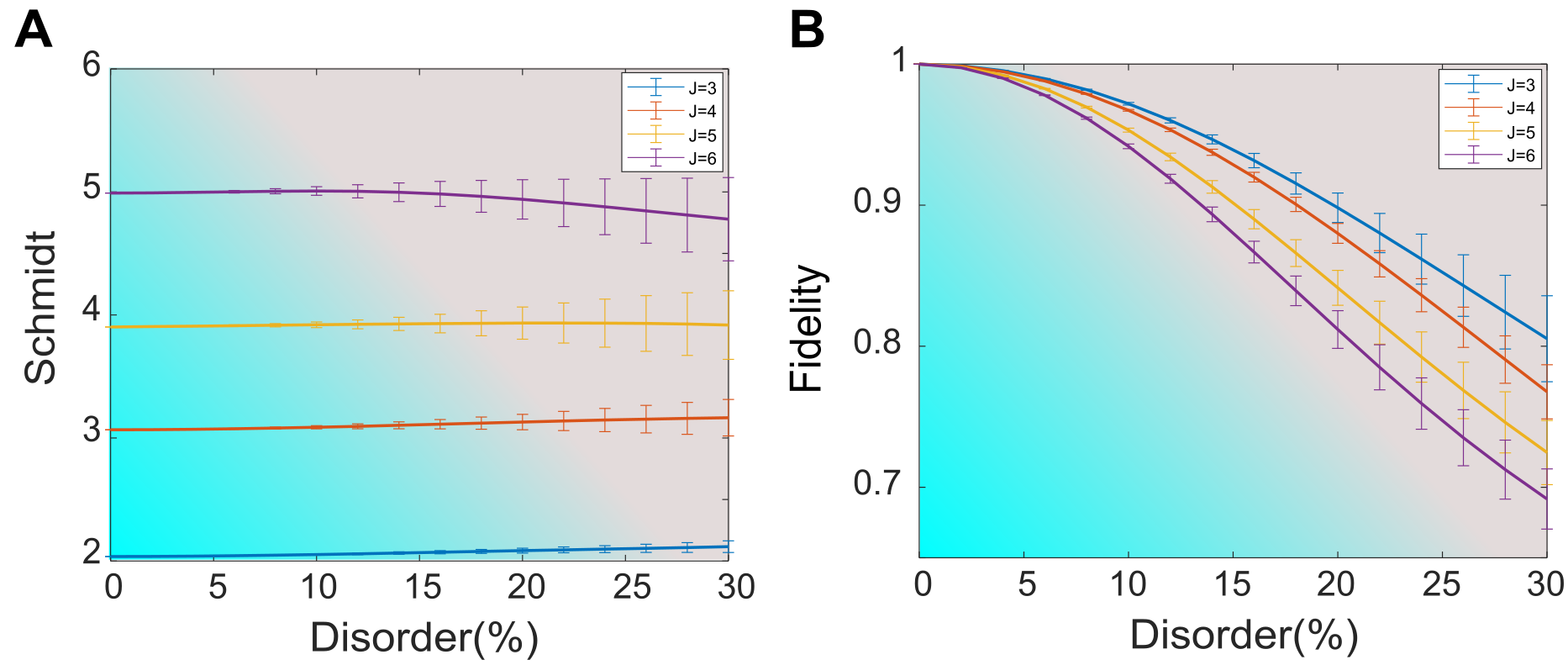}
\caption{Robustness of biphoton entanglement to off\hyp{}diagonal coupling disorder for superlattices with \(J=3,4,5,6\). (A) Schmidt number \(K\) vs.\ disorder: larger \(J\) yields higher mean \(K\); for disorder \(<10\%\), \(K\) is essentially unchanged, while stronger disorder introduces variability that grows with \(J\). (B) State fidelity \(F\) vs.\ the same disorder: all cases start near unity and decrease with increasing disorder; smaller \(J\) retain higher \(F\) under identical perturbations. For disorder \(<10\%\), \(F>0.95\) across all \(J\). Typical nanofabrication variations lie well below \(10\%\).}
\label{FigS.10}
\end{figure}

To further quantify the robustness observed in the correlation maps, we present the Schmidt number \(K\) and the quantum‐state fidelity \(F\). The Schmidt number quantifies the effective dimensionality of biphoton entanglement, with higher values signifying richer modal superpositions. Fidelity measures the overlap between the realized and ideal topological biphoton states \cite{sperling2011schmidt}. Figure \ref{FigS.10}(A) illustrates how $K$ evolves as disorder increases. Superlattices incorporating more sites per unit cell yield higher mean Schmidt numbers, reflecting the participation of additional topological modes in the entanglement. For moderate levels of disorder in the couplings, $<10\%$, the Schmidt number is unaffected. Large levels of disorder result in some variability in $K$, which increases with $J$. 
 
Figure \ref{FigS.10}(B) presents the fidelity response under the same disorder level. All configurations start at near-unity fidelity in the absence of perturbations, but as coupling disorder intensifies, coherence degrades. Importantly, lattices with fewer sites per cell sustain higher fidelity under identical perturbations, whereas more complex superlattices experience a sharper decline. This behavior highlights a trade-off between entanglement dimensionality and disorder tolerance: while increasing unit-cell complexity enriches the biphoton state, it also renders the system more susceptible to decoherence when band gaps narrow. For moderate levels of disorder, $<10\%$, the fidelity remains $>95\%$ for all cases.

\subsection{\label{sec:citeref}Topological and trivial modes}

To benchmark the robustness observed in the simulations, we compare the results for $J$=5 with a comparable lattice supporting four-mode biphoton entanglement, but in this case the entanglement involves two topological modes (red) and two trivial modes(green) [see Fig. \ref{trivial}(A)]. In this case, illustrated in Fig. \ref{trivial}(B), the biphoton correlation is significantly altered by the same amount of disorder, underscoring the susceptibility of trivial modes. To quantify this robustness, we computed a lower-bound Schmidt number and the fidelity of the output state for both systems at various disorder levels. As shown in Fig. \ref{trivial}(C), the mean Schmidt number for the four-mode topological superlattice (red markers) remains relatively flat even at high disorder, whereas it increases sharply in the system involving trivial modes (blue markers). In Fig. \ref{trivial}(D), the topological mode case consistently maintains higher fidelity under moderate to strong disorder, while the mixed case sees a pronounced drop in fidelity. 

\begin{figure}[H]
\centering
\includegraphics[width=1\textwidth]{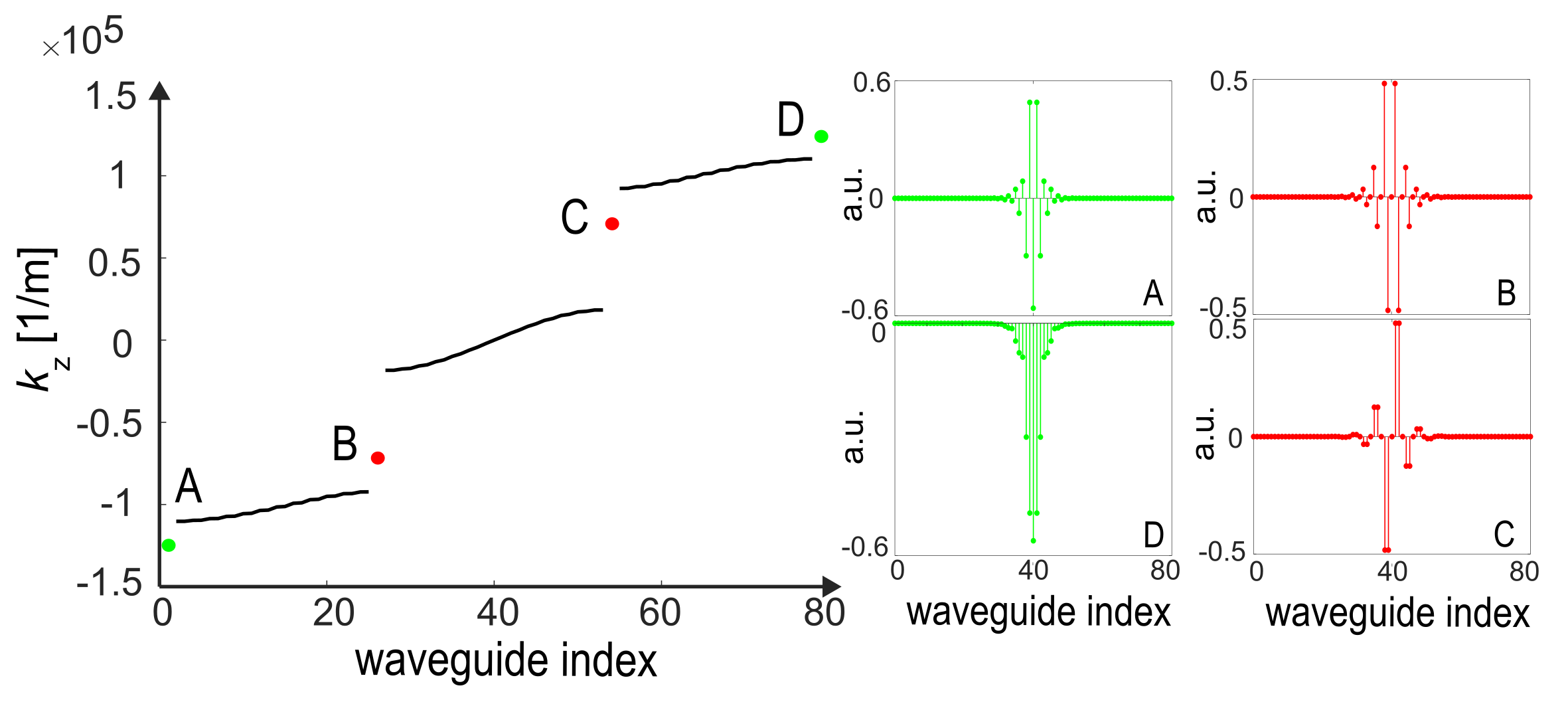}
\caption{Transverse propagation constant $(k_z)$ of the topological(red) and trivial (green) modes. }
\label{trivial}
\end{figure}

\begin{figure}[H]
\centering
\includegraphics[width=1\textwidth]{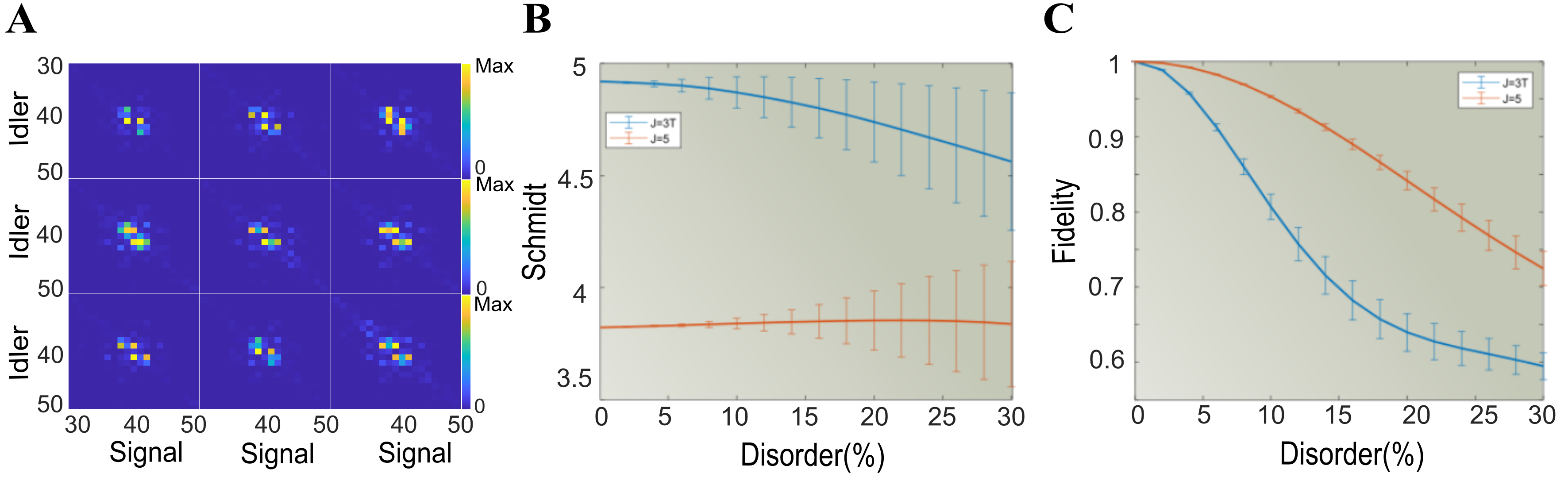}
\caption{(A) Biphoton correlation map at the output of the superlattice for nine random iterations of 10\% off diagonal disorders for the case with both topological and trivial modes (B) Schmidt Number Vs Disorder percentage for J=5 and the case with two topological and two trivial modes (C) Fidelity Vs Disorder percentage for J=5 and the case with two topological and two trivial modes.}
\label{trivial}
\end{figure}

\clearpage

\bibliography{refs}
\bibliographystyle{sciencemag}
\end{document}